\documentclass[sn-mathphys,Numbered]{sn-jnl}% Math and Physical Sciences Reference Style

\usepackage{bm}% bold math
\usepackage{amssymb}
\usepackage{graphicx}%
\usepackage{multirow}%
\usepackage{amsmath,amssymb,amsfonts}%
\usepackage{amsthm}%
\usepackage{mathrsfs}%
\usepackage{appendix}%
\usepackage{xcolor}%
\usepackage{textcomp}%
\usepackage{manyfoot}%
\usepackage{booktabs}%
\usepackage{algorithm}%
\usepackage{algorithmicx}%
\usepackage{algpseudocode}%
\usepackage{listings}%
\usepackage{a4wide}

\newcommand{\be}{\begin{equation}}
\newcommand{\ee}{\end{equation}}
\newcommand{\bea}{\begin{eqnarray}}
\newcommand{\eea}{\end{eqnarray}}
\newcommand{\lan}{\langle}
\newcommand{\ran}{\rangle}
\renewcommand{\slash}{/ \!\!\!\!\,}

\begin{document}

\title{Interactions of $\omega$ Mesons in Nuclear Matter and with Nuclei}

\author{Horst Lenske$^{1}${\footnote{Electronic address: horst.lenske@theo.physik.uni-giessen.de}}}

\affil[1]{ \orgdiv{Institut f\"{u}r Theoretische Physik}, \orgname{Justus-Liebig-Universit\"at Gie\ss en},
\orgaddress{D-35392 Gie\ss en, Germany }}

%-------------------------------------------------------------------------------------
\abstract{
In--medium interactions of $\omega$--mesons in infinite nuclear matter and finite nuclei are
investigated in a microscopic approach, focused on the  particle--hole excitations of the medium involving nucleonic $NN^{-1}$ and $N^*N^{-1}$
modes, where $N^*$ denotes a nucleon resonance. The nuclear polarization tensors include relativistic mean--field dynamics by self--consistent scalar and vector fields. The resulting self--energies are transmitted to finite nuclei in local density approximation. Real and imaginary parts of longitudinal and transversal self--energies are discussed. The relation of the present approach to meson cloud models is addressed and an ambiguity is pointed out. Applications to recent data on the in--medium width of $\omega$ mesons scattered on a Niobium target serve to determine unknown $N^*N\omega$ in-medium coupling constants. The data are well described by $N^*N^{-1}$ self--energies containing S--wave and P--wave $N^*$ resonances. Exploratory investigations, however, show that the spectroscopic composition of self--energies  depends crucially on the near--threshold properties of the width which at present is known only within large error bars. The calculations predict the prevalence of transversal self--energies, implying that vector current conservation is still maintained in the nuclear medium by slightly more than 90\%. Schr\"{o}dinger--equivalent potentials are derived and scattering lengths and effective range parameters are extracted for the longitudinal and transversal channel. Longitudinal and transversal spectral distributions are discussed and the dependencies on momentum and nuclear density are investigated. Schr\"{o}dinger--type $\omega+{}^{93}$Nb optical potentials are constructed. Low--energy parameters are determined, are used to study the pole structure of the S--matrix at threshold. The effective range expansion of the omega--nucleus K--matrix led to a $\omega +{}^{93}$Nb bound states with binding energy $Re(\varepsilon_B)=-448$~keV but of width $\Gamma_B=4445$~keV.
}
%\end{abstract}
%-------------------------------------------------------------------------------------
\date{\today}
\maketitle

\section{Introduction}

An omega--meson in encounter with a nuclear target $A$ will experience various kinds of interactions driven largely by energy-- and momentum--dependent dispersive self--energies. The vector meson self--energy problem has been studied intensively for many years, especially in the context of signals from highly compressed hot matter as reviewed in detail some time ago by Rapp and Wambach \cite{Rapp:1999ej}.
In--medium dynamics of vector mesons was studied theoretically under various aspects from QCD sum rules, e .g
\cite{Margvelashvili:1987tk,Braun:1988qv} where the work of Hatsuda and Lee \cite{Hatsuda:1991ez} was especially influential, to spontaneous chiral symmetry breaking \cite{Harada:2016uca}, and to a broad
spectrum of phenomenological approaches. The dominating interaction scenario for vector mesons in vacuum and in matter
is given by excitations of polarization modes involving both other mesons and baryons. The theory of vector--meson
polarization modes has been addressed under various aspect from meson decay physics in free space, e.g.
\cite{Kaymakcalan:1983qq,Margvelashvili:1987tk,Lublinsky:1996yf}, to frequent studies of in--medium spectral functions
of rho-- and omega--mesons, e.g.
\cite{Celenza:1991ff,Caillon:1995ci,Peters:1997va,Muehlich:2006nn,Shao:2009zzb,Ramos:2013mda,Cabrera:2013zga,Das:2019vhr}. A detailed study of in-medium properties of the omega--meson was performed in \cite{Weil:2012qh} by exploring the $\omega\to \pi^0\gamma$ decay channel in a transport-theoretical approach. A strong motivation for in--medium
investigations is the search for signals of restoration of chiral symmetry as e.g. in \cite{Rapp:2009yu}.

Only lately, the theoretical investigations, typically relying on empirical input, could be evaluated against data for meson--nucleus interaction obtained under on--shell conditions. A comprehensive presentation of the recent experimental
status of meson research on nuclear targets is found in the review article by Metag et al. \cite{Metag:2017yuh}, addressing the demanding challenges of meson-nucleus experiments and their interpretation. By obvious reasons, experimental studies of short-lived particles like the omega--meson ($c\tau=22.73$~fm or $\tau\simeq 7.58\cdot10^{-23}$sec) are a demanding task. A successful method is \textit{in situ} production of the meson on a target and subsequent reconstruction of the spectral distribution by means of the decay channels. Such an approach was used in a photo--production experiment at the ELSA electron accelerator at Bonn, producing omega (and eta') mesons
on a $^{93}$Nb nucleus by incident photon energies of 1.2 to 2.9~GeV  by Kotulla et al. \cite{Kotulla:2008pjl} and Friedrich et al. \cite{Friedrich:2016cms} where the latter work includes also data from the earlier experiment. The CBELSA/TAPS detector system was used to identify the initially produced mesons through the $\omega \to \pi^0\gamma \to 3\gamma$ decay channel. Transparency ratios were extracted over a wide range of incident
energies and used to derive in-medium meson widths by a high--energy eikonal
approximation. The absorptive interactions were assumed to depend on density by a simple scaling law $\Gamma^{(exp)}_{\omega A}(p_{Lab},\rho)\simeq \Gamma^{(exp)}_0(p_{Lab})\frac{\rho}{\rho_{sat}}$, implying
separability of energy and density dependence.

\begin{figure}[h]
\centering
%\sidecaption
\includegraphics[width=13cm,clip]{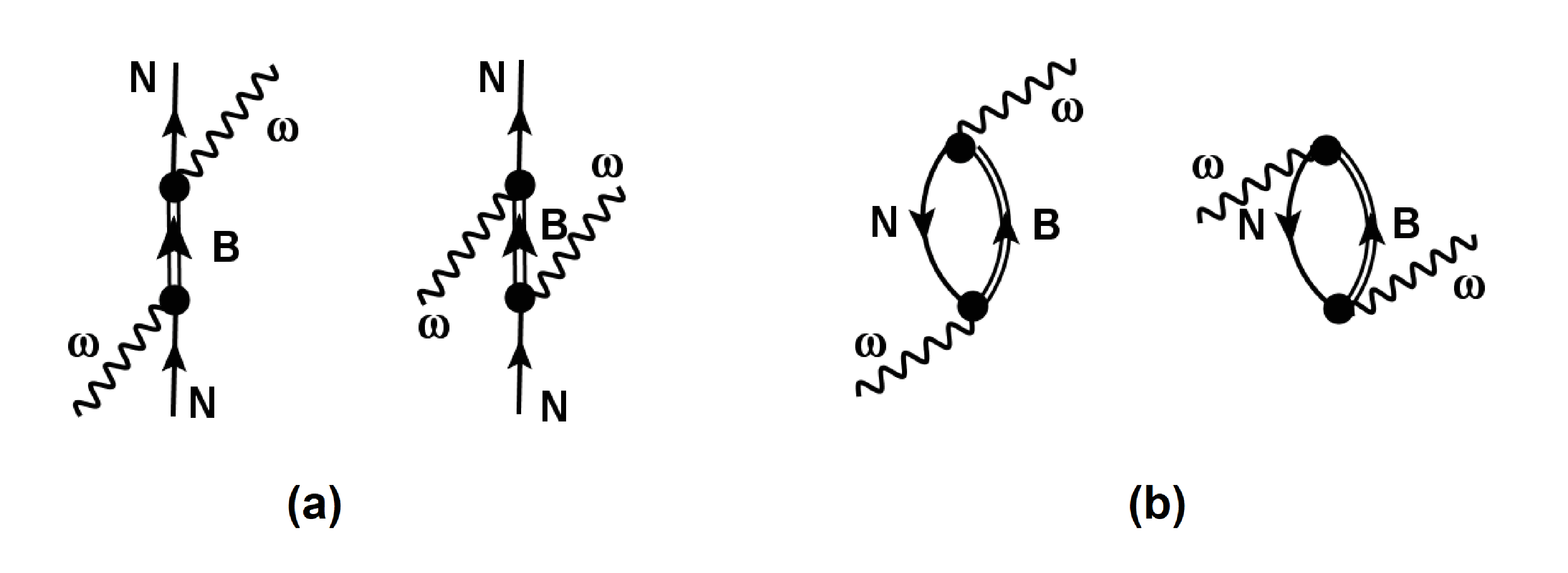}
\caption{The omega--nucleon scattering amplitudes (a) and the corresponding particle--hole type diagrams (b) are depicted. In both cases s--and u--channel diagrams are shown. Nucleons in (a) and hole states in (b) are denoted by $N$. The intermediate $B=N,N^*$ states in (a), which become particle states in (b), are indicated by double lines. }
\label{fig:Graphs}
\end{figure}

In this work, the experimentally constrained in--medium widths of Ref. \cite{Friedrich:2016cms} -- available for energies from threshold to about 2~GeV -- will be used as input in a microscopic approach, serving to determine the model parameters. As depicted diagrammatically in Fig.\ref{fig:Graphs}, the model is focused on dynamical $\omega+A$ self--energies described by nuclear polarization tensors describing particle--hole excitations of the target
\be\label{eq:PTens}
\Pi^{\mu\nu}_{\alpha\beta}=-tr_A\lan A|\widehat{\Gamma}^\mu_{\beta}\mathcal{G}_A\widehat{\Gamma}^\nu_{\alpha}|A\ran .
\ee
They are given by the ground state expectation value of the transition operators $\widehat{\Gamma}^\nu_\alpha$ and the many--body nuclear Green function $\mathcal{G}_A$. $\nu$ is a Dirac-index and, as discussed below, $\alpha$ denotes the type of operator structure. Summations over spin and isospin of the nuclear constituents are indicated by $tr_A$.

The particle--hole type diagrams obtained large attention in early theoretical studies, e.g. \cite{Celenza:1991ff} for the omega--meson and \cite{Peters:1997va} for the rho--meson. In later studies as e.g. \cite{Klingl:1997kf,Lutz:2001mi,Muehlich:2006nn,Rapp:2009yu,Ramos:2013mda} the focus shifted towards in--medium meson loop self--energies, affecting directly the decay channels of the incoming meson. In the following, self--energies from the decay of the omega--meson into other mesons are taken into account on the level of the free space width only. As discussed in appendix \ref{app:Mapping} the (sub--leading) in--medium meson loop diagrams are in fact overlapping partially with the (leading) particle--hole self--energies of Fig. \ref{fig:Graphs}. That leads to an ambiguity for in--medium self--energies with the potential danger of overcounting if both effects are included straight forwardly in a phenomenological approach. In the following, the coupling constants of the omega to the baryon particle--nucleon hole excitations will be determined by fit to the data, thus including to some extent the in--medium modifications of the meson cloud self--energies implicitly.

The primary goal of this work is to study features of $\omega$ meson interaction in cold nuclear matter by the leading order approach based on self--energies determined by the polarization tensors of Eq.\eqref{eq:PTens}. For in--medium studies the widely used practice of fixing coupling constants by the $N^*\to N\omega$ partial decay widths, being used successfully for reactions on the free nucleon, lacks clear justification. Here, the unknown in--medium coupling constants $g_{N^*N\omega}$ will be derived directly by a fit to the $\omega + ^{93}$Nb data of Friedrich et al. \cite{Friedrich:2016cms} on the widths at central density. Nuclei like $^{93}$Nb have a large enough volume for developing an extended region of (almost) constant density as will be seen below in Fig.\ref{fig:gsdNB}. Hence, in a photo--production reaction most of the meson--nucleus interactions occur at densities close to the saturation density of nuclear matter. The results may serve to understand interactions of (isoscalar) mesons in nuclear matter in mechanical and thermodynamical equilibrium, manifestly different from the conditions in central heavy ion collisions.

The theoretical model is introduced in section \ref{sec:Theory}. The approach is based on a description in terms of response functions in infinite asymmetric nuclear matter. The resulting spectral distributions are applied in local density approximation (LDA) to finite nuclei. Recently, a similar approach was applied successfully in the description of resonance production in peripheral charge exchange heavy ion reactions \cite{Rodriguez-Sanchez:2020hfh,Rodriguez-Sanchez:2021zik}.
Applications to the data and the spectroscopic structure of self--energies as well as exploratory studies are discussed in section \ref{sec:Applic}. Longitudinal and transversal self--energies, derived from P--wave and S--wave self--energies, respectively, and spectral distributions are discussed. Examples of dispersive omega self--energies in a finite nucleus, Schr\"{o}dinger-type potentials, low--energy parameters, and indications for a bound state in $^{93}$Nb are explored in section \ref{sec:FiniteNucleus}. In section \ref{sec:SumOut} conclusions are drawn and an outlook to open questions and future work is given. Background material on the theoretical formulation and methods are found in the appendix.

\section{Polarization Tensors and Self--Energies}\label{sec:Theory}
\subsection{General Aspects of Nuclear Response Functions}

As indicated in Fig.\ref{fig:Graphs} and elucidated further in this section, $\omega$--nucleus interactions are described by dispersive self--energies resulting from the absorption of the incoming meson on a nucleon. The nucleon is scattered out of the Fermi sea into either a $NN^{-1}$ or a $N^*N^{-1}$ particle--hole configuration, where the particle state can be either a nucleon $N$ or a nucleon resonance, $N^*$. The approach is a largely updated and extended version of our previous works in \cite{Peters:1997va,Peters:1998hm,Peters:1998mb}.

In momentum representation the polarization tensors describing excitations of $BN^{-1}$ modes, $B=N,N^*$, are given in leading order as
\be
\Pi^{\mu\nu}_{NB}(q)=i(-i)^2
\int \frac{d^4k}{(2\pi)^4}
Tr_s\left(\Gamma^{\mu}_{NB} G_N(k|k_{F})\Gamma^{\nu}_{NB} G_B(q+k|k_{F})+(q\leftrightarrow -q)  \right).
\ee
The trace has to be taken over spin projections.
The set of Fermi--momenta characterizing the occupancies in the ground state is denoted by $k_{F}=\{k_{F_p},k_{F_n} \}$. $G_{N,B}$ are in--medium single particle propagators, $N\in \{p,n \}$ and $B\in \{N,N^* \}$. The Fermi--momenta are defined in terms of the proton ($p$) and neutron ($n$) densities by $k_{F_q}=(3\pi^2\rho_q)^{\frac{1}{3}}$, $q=p,n$.

Performing the energy--integration we arrive at $k_0=E^*_N(\mathbf{k}^2)=\sqrt{\mathbf{k}^2+M^{*2}_N}$ and the in--medium four--momentum $k^{*\mu}=k^\mu+U^\mu_N(\rho_M)$, including effective in--medium masses $M^*_B(\rho_M)$ and mean--field vector self--energies $U^\mu_N(\rho_M)$, being functionals of the (invariant) nucleon number density $\rho_M=\{\rho_p,\rho_n\}$ of the background medium. Inserting RMF dynamics the polarization propagators become
\be
\Pi^{\mu\nu}_{NB}(w,\mathbf{q})=\int \frac{d^3k}{(2\pi)^3}\frac{\theta(k^2_{F_N}-\mathbf{k}^2)\theta(\mathbf{(k+q)}^2-k^2_{F_B})}{2E^*_N(\mathbf{k})}
Tr_s\left(\Gamma^\mu(\slash{k}^*+M^*_N)\Gamma^{\nu}G_B(\mathbf{k}+\mathbf{q}|k_F) \right).
\ee
Since interactions of the $\omega$--meson with a nuclear medium are of isoscalar character, only two kinds of vertex operators are relevant:
\bea\label{eq:SigmaOp}
&&\Gamma^\mu_{NB}= \gamma^\mu \quad\quad \textrm{if B=N,N$^*$ is a positive parity state}, \\
&&\Gamma^\mu_{NB}= \gamma_5\gamma^\mu \quad \textrm{if B=N$^*$ is a negative parity state},
\eea
constraining transitions to isospin $I=\frac{1}{2}$ resonances.
We neglect relativistic rank--2 tensor interactions, known to be negligible in elementary omega--nucleon interactions. In the following, we consider $NN^{-1}$ and $N^*N^{-1}$ polarization tensors, including  $S_{11}$, $S_{13}$ resonances of negative parity and $P_{11}$, $P_{13}$ resonances of positive parity with  masses up to 2.3~GeV. Masses and widths are taken from the literature \cite{PDG:2022}, while couplings constants are treated as free parameters to be derived from data.

\subsection{Longitudinal and Transversal Self--Energies}
The dispersive self--energies are directly proportional to the nuclear polarization tensors. In asymmetric infinite matter $A$ one finds
\be\label{eq:SelfE}
\mathcal{S}^{\mu\nu}_{\omega A}(w,\mathbf{q})=\sum_{N;B}g^2_{\omega NB}\Pi^{\mu\nu}_{NB}(w,\mathbf{q}),
\ee
depending besides on energy and momentum also on the total density of the medium and, to a lesser extent, also on the proton and neutron content separately through their Fermi--momenta.

As discussed in the Appendix, by means of the orthogonal projectors $P^{\mu\nu}_{L/T}$, obeying $P_{L}+P_{T}=1$, $P^2_{L/T}=P_{L/T}$, and $P_L\cdot P_T=P_T\cdot P_L=0$,
the self--energy tensor is decomposed into longitudinal (L) and transversal (T) components:
\be
\mathcal{S}^{\mu\nu}(w,\mathbf{q})=\left(P^{\mu\nu}_{L}+P^{\mu\nu}_{T} \right)^2\mathcal{S}^{\mu\nu}(w,\mathbf{q})=
P^{\mu\nu}_{L}\Sigma_L(w,\mathbf{q})+P^{\mu\nu}_{T}\Sigma_T(w,\mathbf{q})
\ee
where the (scalar) longitudinal and transversal reduced self--energies are defined by inversion
\be
\Sigma_{L/T}(w,\mathbf{q})=P^{\mu\nu}_{L/T}\mathcal{S}_{\mu\nu}(w,\mathbf{q})
\ee
From Eq.\eqref{eq:SelfE} it is seen that the projections are acting in fact on the polarization tensors, implying the corresponding decomposition
\be
P^{(NB)}_{L/T}(w,\mathbf{q})=P^{\mu\nu}_{L/T}\Pi_{\mu\nu,NB}(w,\mathbf{q}).
\ee
Hence, we obtain
\be
\Sigma_{L/T}(w,\mathbf{q})=\sum_{N=p,n;B}g^2_{BN\omega}P^{(NB)}_{L/T}(w,\mathbf{q}).
\ee
Explicit formulas are found in the Appendix.

\section{Application to $\omega+{}^{93}$Nb}\label{sec:Applic}

\subsection{Relating the Model to Data}

In the energy region covered by the data, $p_{Lab}\leq 2.6$~GeV/c ($T_{Lab}\leq 1.933$~GeV), polarization self--energies given by excitations of P--wave and S--wave resonances are the by far prevailing omega--nucleus interaction modes. Since the S--wave modes lead to self--energies and widths which are non--vanishing at the kinematical threshold, they are especially important for exploring the low--energy properties of omega--nucleus interactions.

As discussed in the Appendix, the P--wave contributions are of pure transversal character while S--wave modes produce both longitudinal and transversal self--energies. In our case, the total longitudinal self--energy is fully determined by the longitudinal S--wave component:
\be
\Sigma_L(w,\mathbf{q})\equiv  \Sigma^{(S)}_L(w,\mathbf{q})
\ee
while the total transversal self--energy contains P--wave and S--wave components
\be\label{eq:SigmaT}
\Sigma_T(w,\mathbf{q})= \Sigma^{(S)}_T(w,\mathbf{q})+ \Sigma^{(P)}_T(w,\mathbf{q})
\ee
Accordingly, transversal and longitudinal partial widths are obtained:
\be
\Gamma_{L/T}(\mathbf{q})=-\frac{1}{m_\omega}Im\left(\Sigma_{L/T}(E(\mathbf{q}),\mathbf{q})  \right)
\ee
where on--shell kinematics are used.

The available data do not provide separate information on the amount of transversal and longitudinal self--energies. This caveat poses the challenge to theory how to extract information on the two kinds of self--energies from a single set of data. In other words, the data implicitly contain -- in principle unknown -- superpositions of longitudinal and transversal self--energies.

\subsection{Details of the Numerical Calculations}

As discussed in App.\ref{sapp:PolProp} the polarization tensors are evaluated in relativistic mean--field (RMF) approximation. This means to include static scalar and vector self--energies of isoscalar and isovector character for nucleons and resonances likewise. The mean--field interactions lead to effective masses $M^*_{B}(\rho_M)$  and vector potentials $U^\mu_{B}(\rho_M)$, both containing isoscalar and isovector interactions, depending on the isoscalar and isovector nuclear ground state densities. In the nuclear rest frame, the vector potentials are chosen as purely time--like, $U^\mu_{B}(\rho_M)=\delta^{\mu 0}U_B(\rho_M)$. The RMF part of the theory is based on the Giessen Dirac-Brueckner-Hartree-Fock (DBHF) approach \cite{Hofmann:2000vz,Lenske:2004vm,Adamian:2021gnm}. The proton and neutron scalar and vector ground state densities and RMF potentials are derived self--consistently as discussed in \cite{Hofmann:2000vz}. The mean--field part involves scalar $f^{(I)}_{BBs}$ and vector $f^{(I)}_{BBv}$ coupling constants for isoscalar ($I=0$) and isovector $(I=1)$ interactions. In a meson exchange description, the scalar fields are given by $\sigma$ ($I=0$) and $\delta/a_0(980)$ ($I=1$) exchange, the vector fields are described by $\omega$ ($I=0$) and $\rho$ ($I=1$) exchange \cite{Hofmann:2000vz,Lenske:2004vm}. The meson fields are the ground state expectation values of the corresponding field operators intrinsic to the nuclear system. Note that the mean--fields are highly virtual objects of static character and especially the omega field must not to be confused with the physical on--shell omega reacting with the nucleus.

Proton and neutron mean--field dynamics are well known which, however, is not the case for resonances. Thus, the question arises how to fix the $B=N^*$ mean--field coupling constants which are unknown and are, indeed, empirically hardly accessible. They were chosen equal to the nucleon coupling constants,
$f^{(I)}_{N^*N^*a}\equiv f^{(I)}_{NNa}$ for scalar ($a=s$) and vector ($a=v$) interactions. By this -- admittedly somewhat arbitrary -- choice, an overabundance of free parameters is avoided. A practically convenient side effect is that in the particle--hole propagators the vector potentials of nucleon holes and resonances particle states cancel mutually. The action of scalar potentials, however, remains visible through higher order effects of effective masses. Mean--field interactions of the omega--meson are neglected.

For the self--energies the problem occurs that the respective in-medium particle--hole coupling constants $g_{NN\omega}$ and $g_{N^*N\omega}$ are unknown. It should be noted that conceptually these coupling constants belong to residual interactions attached to particle--hole vertices which in general are of a different structure than the mean--field Hartree--vertices as is evident from Fermi liquid theory and density functional theory (DFT), respectively. The widely used practice to derive their values from the $N^*\to N\omega$ partial decay widths of baryon resonances in free space is highly questionable for in--medium calculations. As found in covariant DFT, e.g. \cite{Hofmann:2000vz,Lenske:2004vm}, in--medium interaction vertices are functionals of the nucleon field operators. In mean--field approximation they become functions of the ground state density. Typically, their values decrease in magnitude when the density of the surrounding medium increases. Here, the residual interaction couplings are treated as free parameters.

The problem of unknown coupling constants affects on a different level also the practical computation of the polarization tensors. An elaborate full approach would mean to sum the full series of mixed $NN^{-1}$ and $N^*N^{-1}$ loop diagrams, i.e. to derive self--energies in Tamm-Dancoff-Approximation (TDA) or Random--Phase-Approximation (RPA). A full solution of either set of coupled linear systems depends, however, on the knowledge of the interactions and matrix elements connecting the various kinds of  particle--hole configurations. That input is known safely only for the $NN^{-1}$ sector but completely lacking for the resonance channels. In view of these problems, self--energies will be described throughout in lowest order.

In Tab.\ref{tab:Tab1} the spectrum of $N^*$ states, their masses and widths are shown. In addition, also the (covariant) effective masses at saturation density are displayed in units of the corresponding mass in free space. From those numbers, it is evident that $NN^{-1}$ and the lowest $N^*N^{-1}$ excitations are below threshold, thus contributing rarely to the width but possibly affecting the real parts of self--energies.

\begin{table}
  \centering
\begin{tabular}{|c|c|c|c|c|}
  \hline
  Particle & Mass/MeV & total Width/MeV & $g_{N^*N\omega}$& $M^*/M$\\
 \hline
  $^{93}$Nb      & 86.521$\times 10^3$ & -- & --& --\\
  $\omega$       & 782.7               & 8.68& --  & 1\\
  P$_{11}$(940)  & 938.92 & --& 2.076&0.60 \\
  P$_{11}$(1440) & 1386 & 350 & 0.218   &0.72\\
  P$_{11}$(1710) & 1670 & 140 & 2.163    &0.78 \\
  P$_{11}$(1880) & 1880 & 230 & $<10^{-3}$&0.80\\
  P$_{11}$(2100) & 2100 & 260 & $<10^{-3}$&0.82\\
  P$_{11}$(2300) & 2300 & 340 & 1.714&0.84\\
  P$_{13}$(1720) & 1720 & 225 & 0.022&0.78\\
  P$_{13}$(1920) & 1920 & 215 & 0.017&0.80\\
  S$_{11}$(1535) & 1535 & 150 & $< 10^{-3}$&0.76\\
  S$_{11}$(1650) & 1650 & 125 & 0.008&0.77 \\
  S$_{11}$(1895) & 1880 & 120 & 0.006&0.80\\
  S$_{13}$(1520) & 1520 & 110 & 0.003&0.75\\
  S$_{13}$(1700) & 1700 & 200 & 0.011&0.78\\
  S$_{13}$(1875) & 1875 & 200 & 1.554&0.70\\
  S$_{13}$(2120) & 2120 & 300 & 0.008&0.82\\
  \hline
 \end{tabular}
  \caption{Masses and widths, taken from \cite{PDG:2022}, of particles considered in the calculations
  and the $\omega$ coupling constants, derived from the fit to the data of Ref. \cite{Friedrich:2016cms}.
  The self--energies where calculated in local density approximation using self--consistent RMF-Hartree $^{93}$Nb
  ground state densities where the central isoscalar density is $\rho_A=0.140 fm^{-3}$. For completeness,
  also the rest mass of the target nucleus is displayed in the first row. The ratio of the effective in--medium masses
  relative to the free--space masses in the center of $^{93}$Nb are listed in the last column.} \label{tab:Tab1}
\end{table}

The coupling constants derived by a first exploratory separate S-- and P--wave fit to data, explained below, are listed as well in Tab.\ref{tab:Tab1}. Most of the coupling constants are seen to be compatible with zero. Only those $N^*$ states with sizable coupling constants are considered in the final combined P/S--wave analysis.

As seen in Tab.\ref{tab:Tab1} the calculations include $NN^{-1}$ modes and the full set of seven P--wave and seven S--wave resonances of isospin $I=\frac{1}{2}$ and $J^\pi=\frac{1}{2}^\pm,\frac{3}{2}^\pm$  up to 2.3~GeV. Thus, in total 15 coupling constants have to be fixed on, however, only 13 data points. In view of this imbalance, the following fitting strategy is chosen: For the sake of an unbiased description, a first exploratory analysis is performed by separate fits with only P--wave and only S--wave self--energies, respectively. These preparatory calculations serve to identify the self--energy modes of largest importance for the description of the data. As will be seen, two P--wave resonance modes and two S--wave modes are of highest relevance for the widths in the available energy interval. The $NN^{-1}$ modes are minor importance for the imaginary part of the self--energy and width but contribute to the real part of the transversal self--energy.  These five modes are then included in a combined fit leading to the final updated results. The coupling constants are listed in Tab.\ref{tab:Tab1}.

Rather than separating and parameterizing the density dependence here we use the full density dependence of the self--energies as predicted theoretically. As seen in Fig.\ref{fig:gsdNB} below, the $^{93}$Nb RMF calculations predict $\rho_A=0.140 fm^{-3}$, undershooting nuclear saturation density by about 10\%. The proton and neutron central vector densities are $\rho_p(0)=0.063  fm^{-3}$ and $\rho_n(0)=0.077  fm^{-3}$, respectively. The charge asymmetry
factor is $Z/A=41/93\simeq 0.44 \simeq \rho_p(0)/\rho_A$, being used throughout in all self--energy calculations.

\subsection{Explorative Calculations with P--wave and S--wave Self--Energies}

\begin{figure}[h]
\centering
%\sidecaption
\includegraphics[width=7cm,clip]{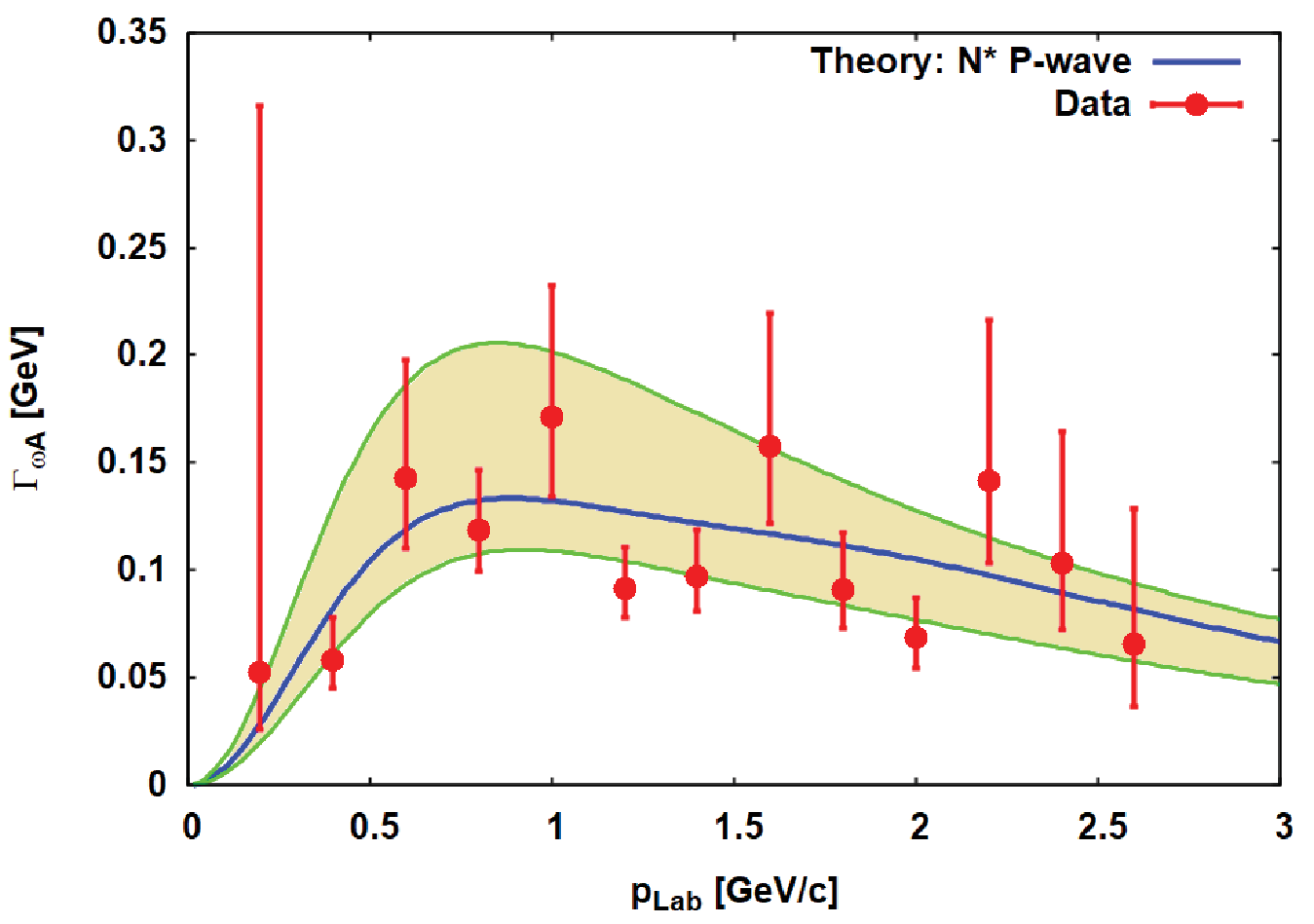}
\includegraphics[width=7cm,clip]{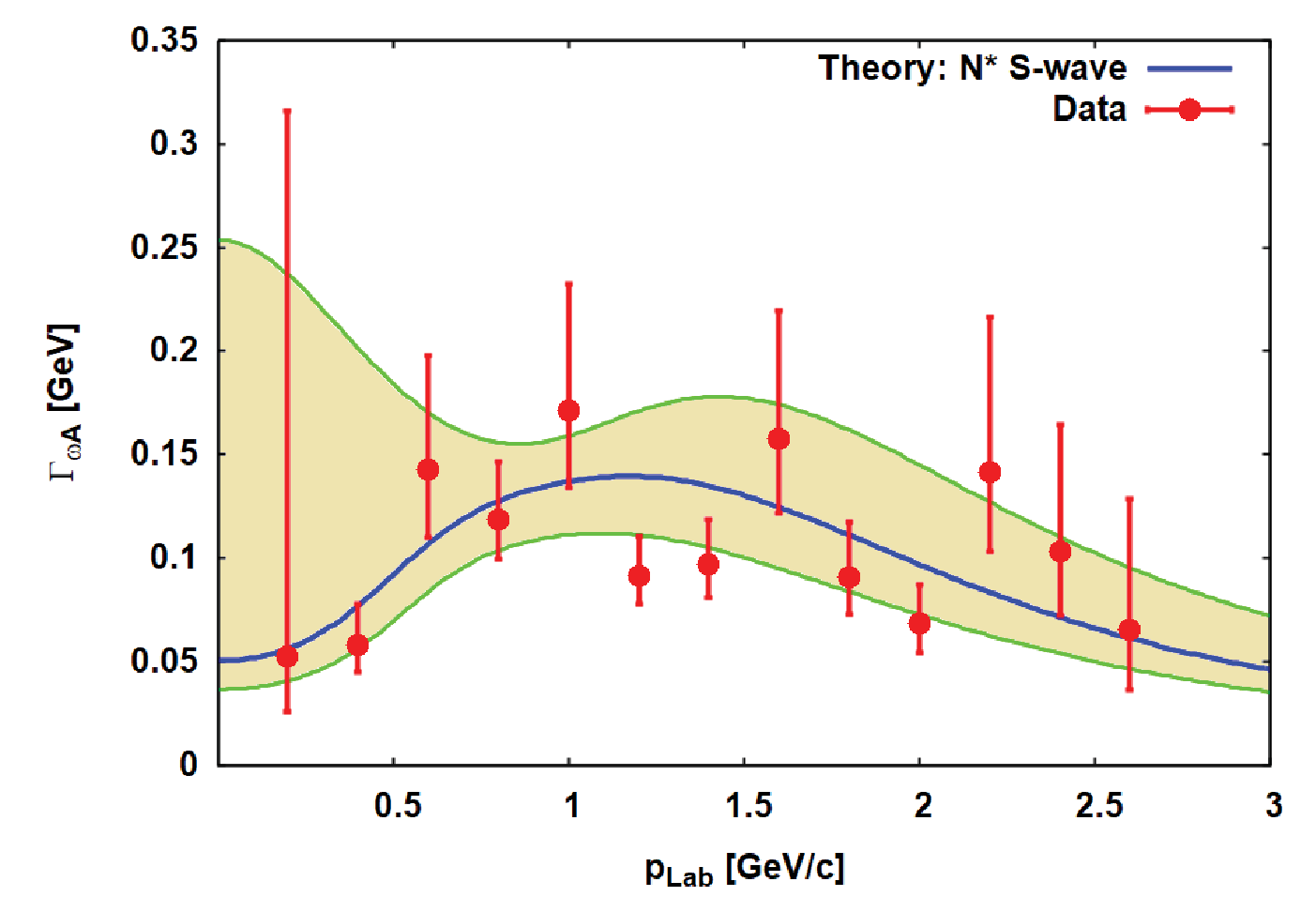}
\includegraphics[width=7cm,clip]{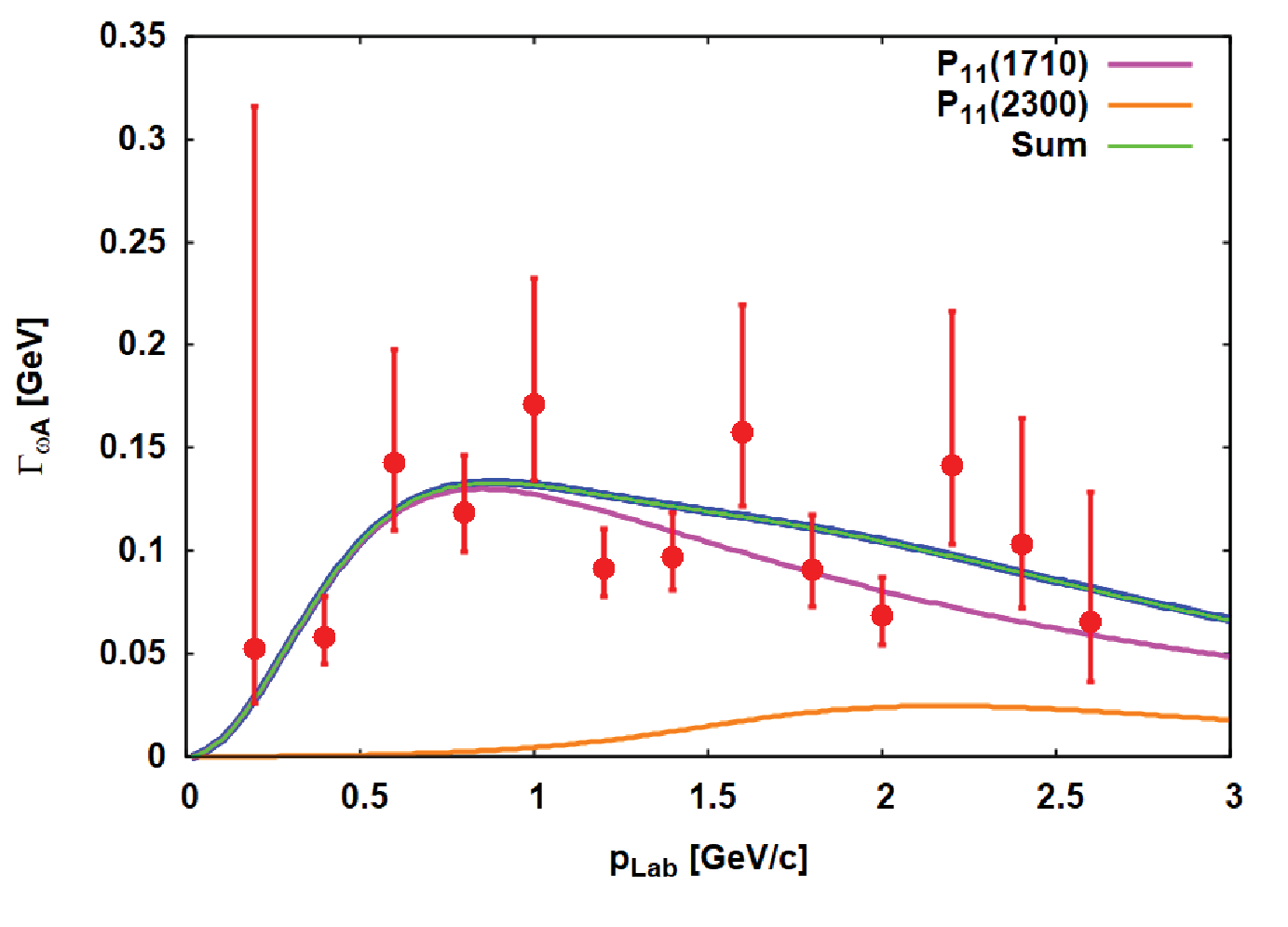}
\includegraphics[width=7cm,clip]{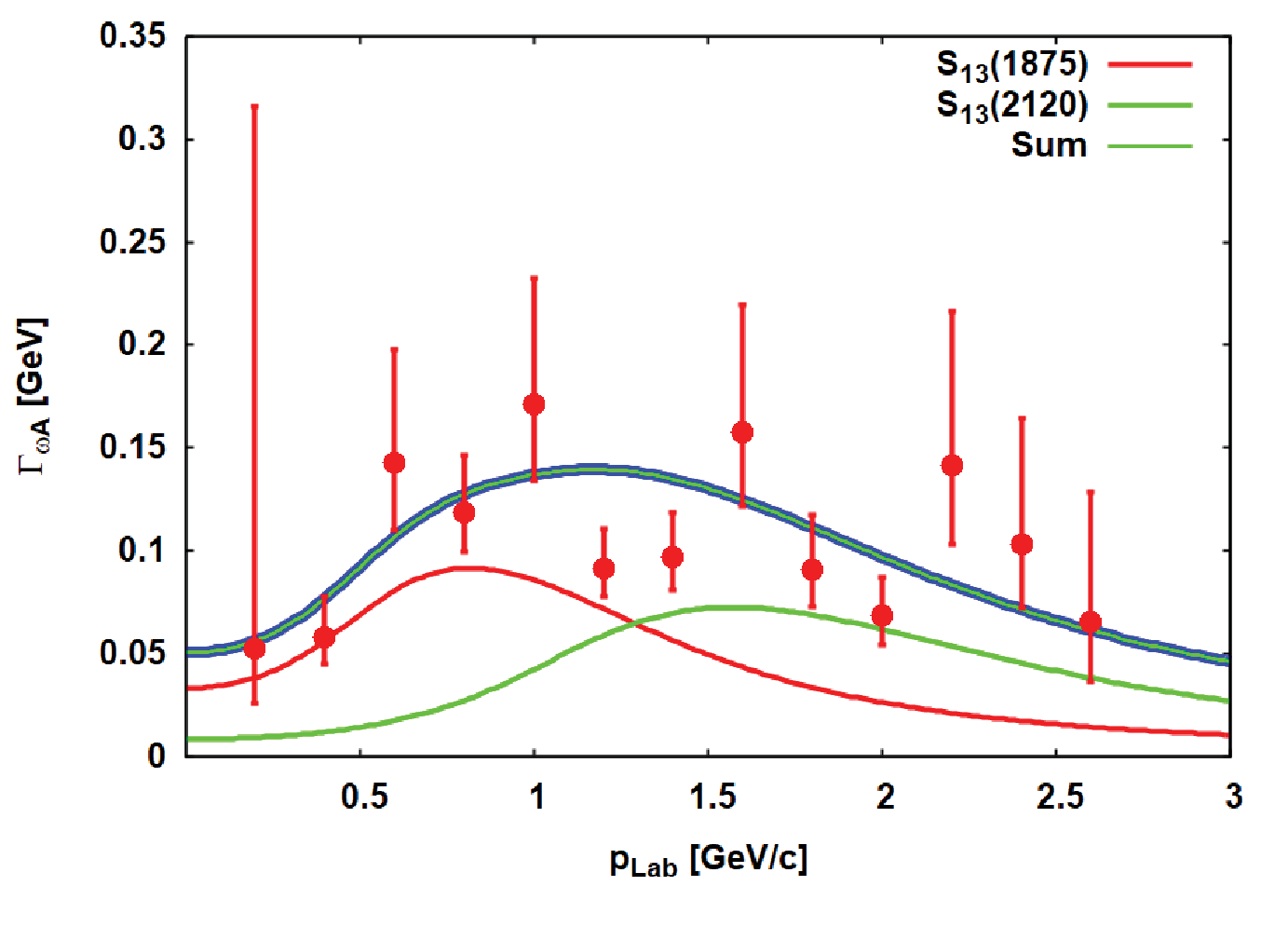}
\caption{The in--medium width of an $\omega$ meson at central density $\rho_A=0.140 fm^{-3}$ of $^{93}$Nb is shown as
 function of the omega momentum $p_{Lab}$.
 Results of the exploratory model calculations assuming artificially only P--wave (left column) and only S--wave self--energies (right column) are compared to the experimentally deduced values of Ref. \cite{Friedrich:2016cms}.
 Uncertainties in the theoretical widths due to the
 experimental errors are indicated by the shaded area (colour online). In the lower figure,  for either case the dominant partial contributions and their respective sums are shown. See text for further discussion.}
\label{fig:Gw_Expl}
\end{figure}

In Fig.\ref{fig:Gw_Expl} the widths obtained from the exploratory P-wave and S--wave separate fits are compared to $\Gamma_{\omega A}(p_{Lab})$ of \cite{Friedrich:2016cms}. The major difference of the two types of calculations is the behaviour close to threshold where the P--wave width approaches zero while the S--wave width remains finite. The large experimental uncertainties, however, induce an outstanding large uncertainty in the low--energy behaviour of the S--wave part. The large S--wave threshold--strength at the upper limit is solely due to an extraordinary enhancement of the response involving $R=S_{11}(1650)$ which is negligible otherwise. The fit to the upper data points imposes in the threshold region a strong constraint on shape and downward slope of the functional form of the width. That translates into a large coupling constant, $g_{RN\omega}\sim 3.97$. Adjusting the width to the median, however, requires a functional form with upward slope which suppresses the $S_{11}(1650)$  self-energy to a negligible level.

The overall S--wave fit to the center values still leads to $\chi^2_S=1.321$ which is comparable to the P--wave result, $\chi^2_P=1.086$. Hence, leaving out the pronounced differences at threshold, these exploratory calculations lead to the conclusion that both kinds of self--energies are compatible with the data and must be taken into account in a full analysis. In the following studies, the sizable uncertainties at threshold will not be considered further but have to be kept in mind as a caveat.

The major issue of these exploratory model calculations was to learn about the spectroscopic details hidden in the width distribution. In the lower row of Fig.\ref{fig:Gw_Expl}, the dominant partial widths of the P--wave and S--wave self--energies are shown. The P--wave width is practically completely given already by the self--energies involving excitations of $N^*=P_{11}(1710)$ and $N^*=P_{11}(2300)$ resonances, plus a minor contribution of $NN^{-1}$ modes. The S--wave case is dominated by two $N^*N^{-1}$ modes with $N^*=S_{13}(1875)$ and$N^*=S_{13}(2120)$. A common feature of the P--wave and S--wave models is that a low--mass and a high--mass resonance are sufficient to reproduce the data. While in the P--wave sector only $J^\pi=\frac{1}{2}^+$ resonances contribute significantly, the S--wave model calculations favor clearly $J^\pi=\frac{3}{2}^-$ resonances.

\subsection{Combined P--Wave and S--wave Self--Energies}\label{ssec:Combined}
The results of the previous section are indicating clearly that the behaviour of the observed width close to threshold is decisive for the determination of the S--wave content of $\omega+A$ self--energies and in turn also affects the amount of P--wave self--energies over the whole energy region. The strategy followed in this section is to keep fixed the $NN^{-1}$ coupling constant to the value of Tab.\ref{tab:Tab1}, while the coupling constants of the self--energies involving $S_{13}(1875)$, $S_{13}(2120)$ and $P_{11}(1710)$, $P_{11}(2300)$, respectively, are readjusted to the data of Ref. \cite{Friedrich:2016cms}.

Two scenarios are investigated: In the \textit{unconstrained scheme} (US), the fit is performed by free variation of the four coupling constants. In the \textit{constrained scheme} (CS), the in--medium total width is chosen as $\Gamma_{thr}=\Gamma_{\omega A}(0)=\Gamma^{(A)}_0+\Gamma_{free}$ with $\Gamma_{free}=8.68$~MeV and imposing an arbitrarily chosen value for the in--medium width $\Gamma^{(A)}_0$.

The coupling constants derived by the US--fit to the central values of the data are shown in Tab.\ref{tab:US_Cpl}. Compared to Tab.\ref{tab:Tab1}, the largest changes appear for the coupling constants of the two S--wave channels. Overall, the $N^*N^{-1}$ coupling constants are of order one. An exception is $S_{13}(2120)$. In the independent S--wave fit the modes including that resonance were essential for a good description of the high energy part of the width. In the combined final fit, however, that energy sector is covered by the P--wave $N^*N^{-1}$--modes with $N^*=P_{11}(2300)$.

As seen in Tab.\ref{tab:US_Cpl}, the final P--wave couplings are uncertain by about 30\%. The uncertainties of the S--wave couplings range from 18\% for $S_{13}(1875)$ to more than 40\% for $S_{13}(2120)$. The smaller uncertainty in the $S_{13}(1875)N\omega$ coupling is the combined result of the close--to--threshold location of $S_{13}(1875)N^{-1}$ excitations and the normative power exerted by the width at threshold on the parameters, thus emphasizing again the importance of precise low--energy data and an appropriate theoretical choice of states.
Since neither the P--wave nor the higher S--wave modes are controlled by these relatively narrow constraints, their coupling constants carry larger uncertainties.
In the following calculations the coupling constants of Tab.\ref{tab:US_Cpl} will be used throughout.

\begin{figure}[h]
\centering
%\sidecaption\overrightarrow{}
\includegraphics[width=7cm,clip]{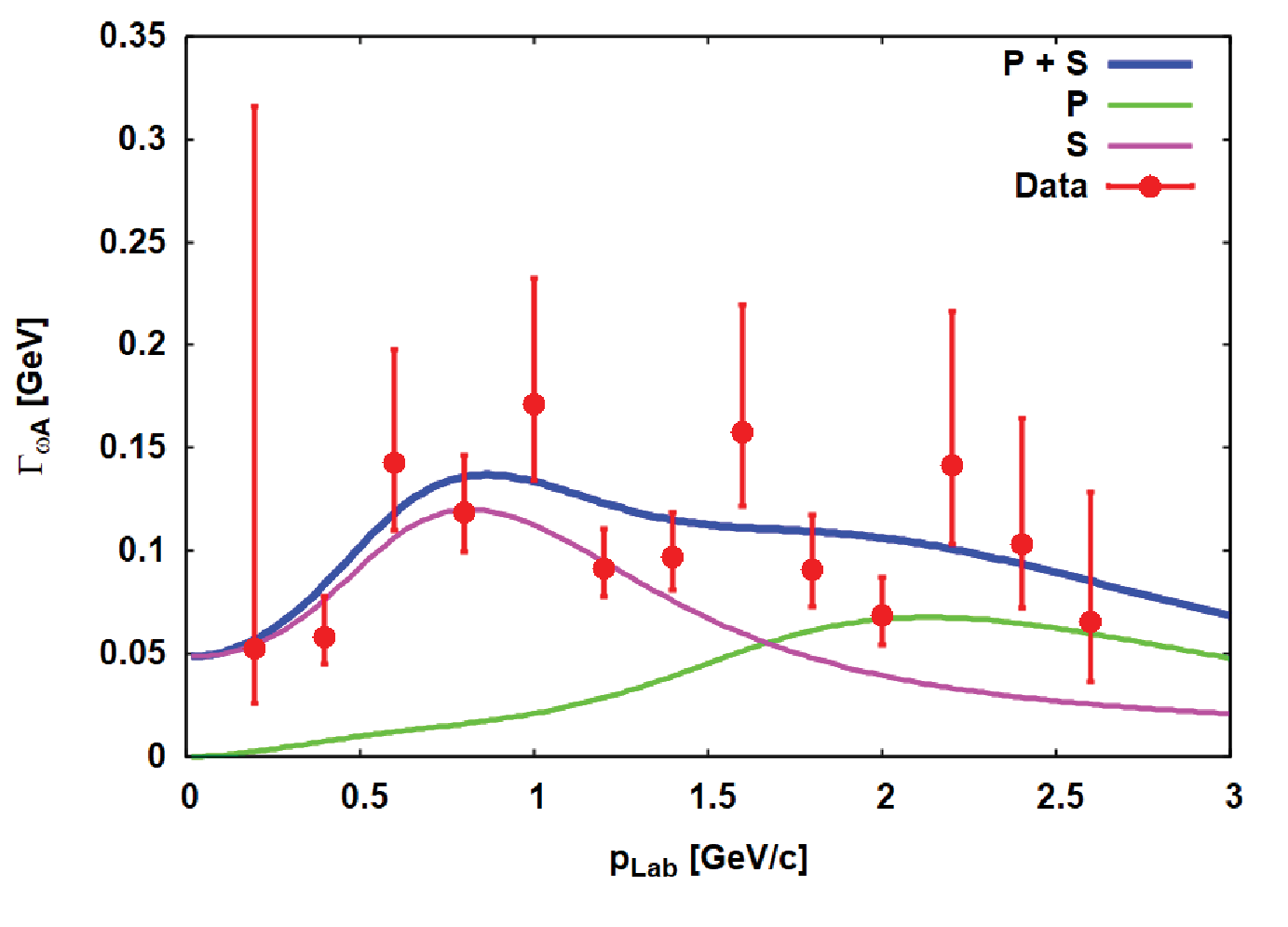}
\includegraphics[width=7cm,clip]{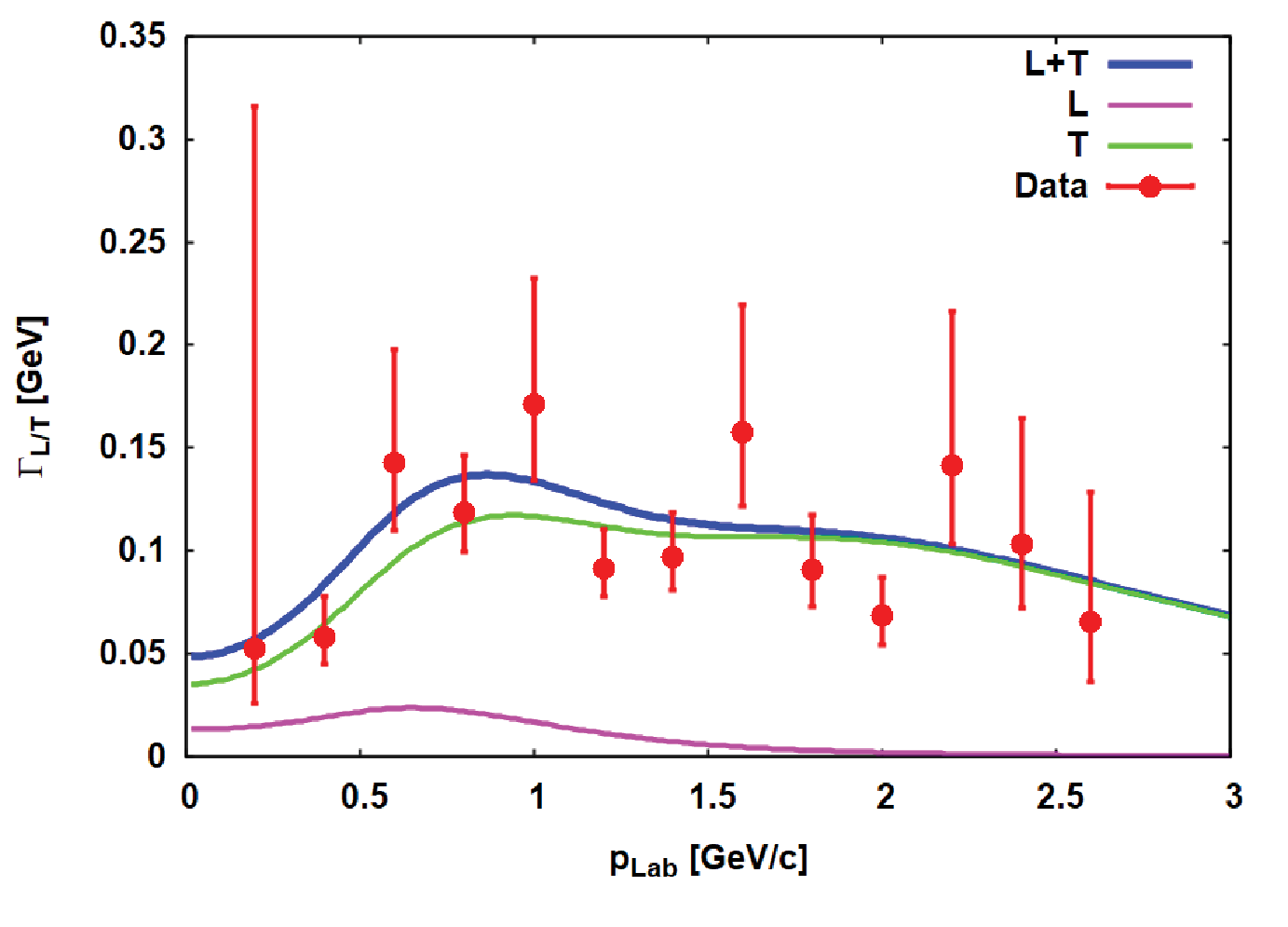}
\includegraphics[width=7cm,clip]{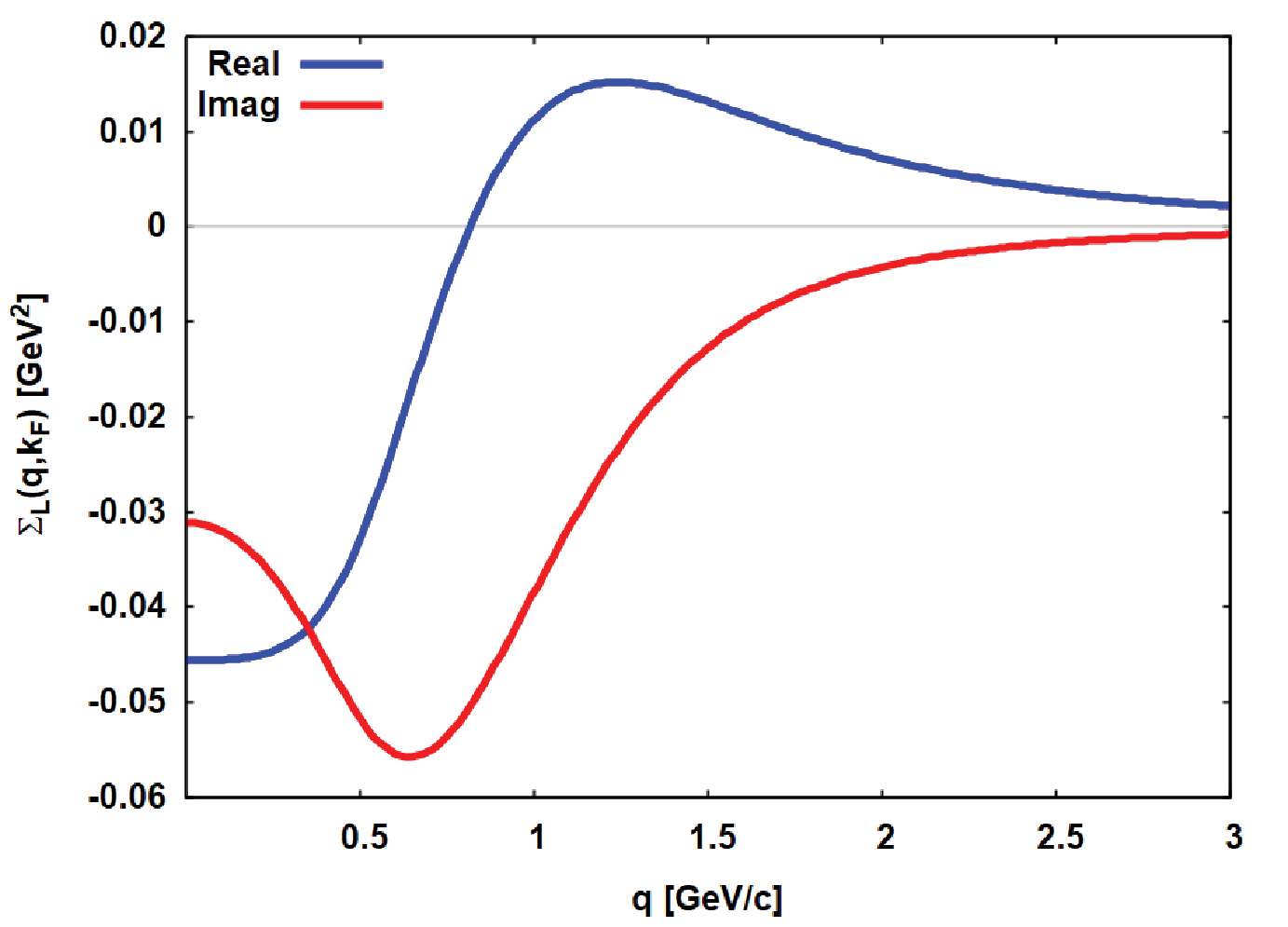}
\includegraphics[width=7cm,clip]{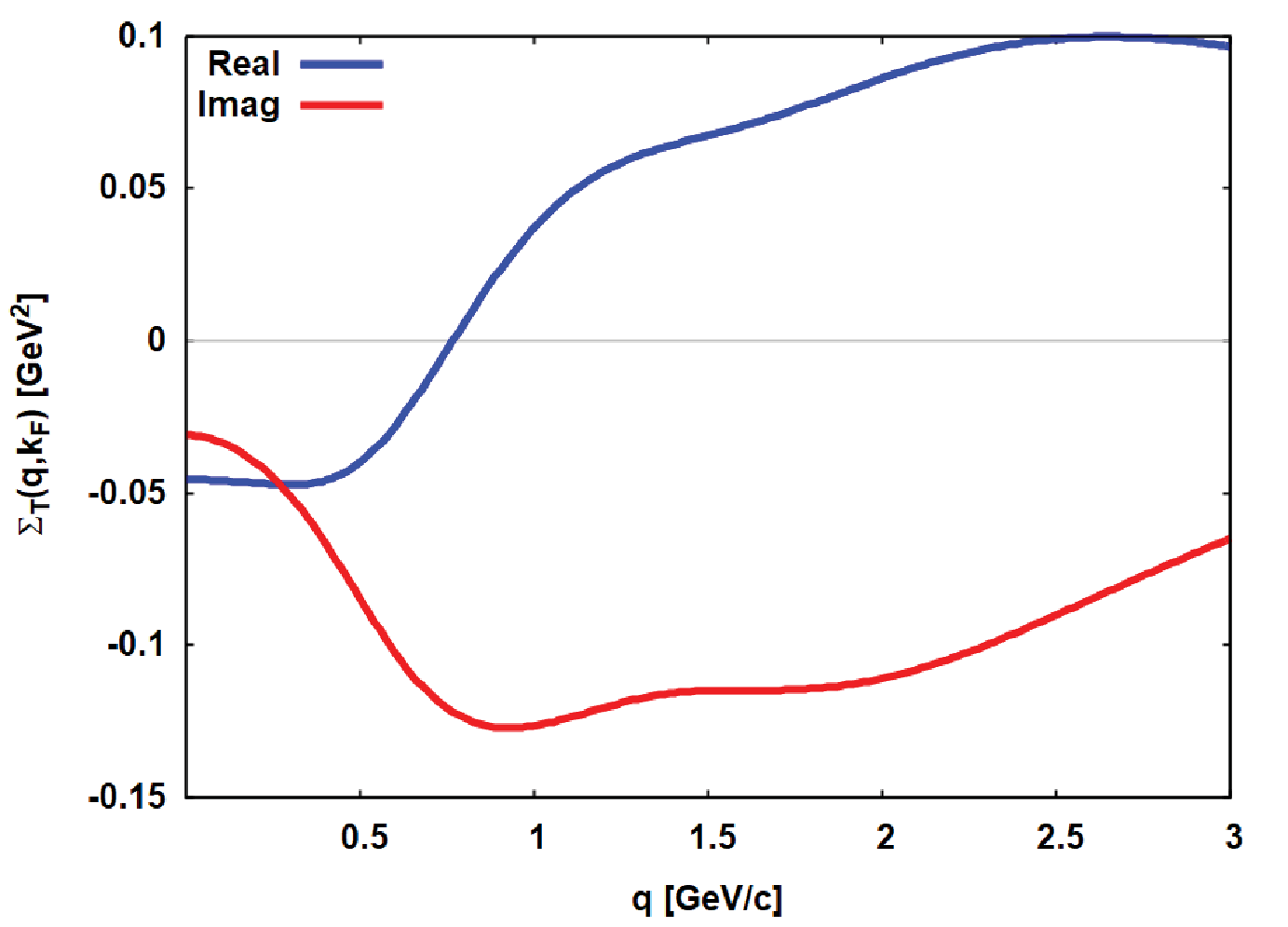}
\caption{Comparison of the theoretical width distributions as functions of momentum to the $\omega +{}^{93}$Nb data of Friedrich et al. \cite{Friedrich:2016cms}. The P--wave (P) and S--wave (S) and longitudinal (L) and transversal (T) partial contributions, shown in the left and the right panel respectively, are evaluated at central density of $^{93}$Nb, $\rho=0.140fm^{-3}$, taking into account the differences of proton and neutron densities. The results were obtained by unconstrained fits to the data, taken again from Ref. \cite{Friedrich:2016cms}. The derived longitudinal (bottom left) and transversal (bottom right) self--energies are displayed in the bottom row.}
\label{fig:GwPSuLT}
\end{figure}

In Fig.\ref{fig:GwPSuLT} the in--medium widths from the US--fit are displayed. The free, unconstrained variation of coupling constants leads to a width at threshold $\Gamma_{thr}=48.41$~MeV, and after subtraction of the free omega width a medium-dependent width $\Gamma^{(A)}_0=39.73$~MeV is obtained. That result is much smaller than the values predicted by other approaches utilizing the processes depicted Fig.\ref{fig:wpirho}. For example, Ramos et al. \cite{Ramos:2013mda} obtained $\Gamma_{\omega A}(0)= 120\pm 10$~MeV while Cabrera and Rapp \cite{Cabrera:2013zga} found in--medium widths as large as $\Gamma_{\omega A}(0)\sim 200$~MeV.

Integrated over the energy interval up to the $p_{Lab}= 3$~GeV/c ($T_{Lab}\simeq 2.32$~GeV), S--wave and P-wave components are found to account for about 57\% and 43\%, respectively, of the total US-yield. These numbers confirm quantitatively that the S--wave modes are substantial for a proper description of the near--threshold region while P--wave self-energies are important for the reproduction of the observed width at higher energies. As seen in Tab.\ref{tab:Tab2}, the S-/P--wave yields correspond to longitudinal and transversal fractions of about 10\% and 90\%, respectively, strongly emphasizing the prevalence of the transversal channel. In other words,  vector current conservation is essentially maintained in the nuclear medium by up to about 90\%.

\begin{table}
  \centering
\begin{tabular}{|c|c|c|c|c|}
  \hline
  Particle & Mass/MeV & total Width/MeV & $g_{N^*N\omega}$& uncertainty in \% \\
 \hline
  P$_{11}$(940)  & 938.92 & --& 2.076 $\pm$ 0.165   & 7.96 \\
  P$_{11}$(1710) & 1670 & 140 & 1.738 $\pm$ 0.588   & 33.80 \\
  P$_{11}$(2300) & 2300 & 340 & 1.795 $\pm$ 0.514   & 28.62 \\
  S$_{13}$(1875) & 1875 & 200 & 1.737 $\pm$ 0.320   & 18.40 \\
  S$_{13}$(2120) & 2120 & 300 & 0.028 $\pm$ 0.012   & 43.55 \\
  \hline
 \end{tabular}
  \caption{Resonances taken into account finally in the combined P-- and S--wave unconstrained fit to the data of \cite{Friedrich:2016cms}. The derived  coupling constants and absolute uncertainties are shown in the third column. The relative uncertainties are displayed in the last column.}\label{tab:US_Cpl}
\end{table}

\begin{figure}[h]
\centering
%\sidecaption\overrightarrow{}
\includegraphics[width=7cm,clip]{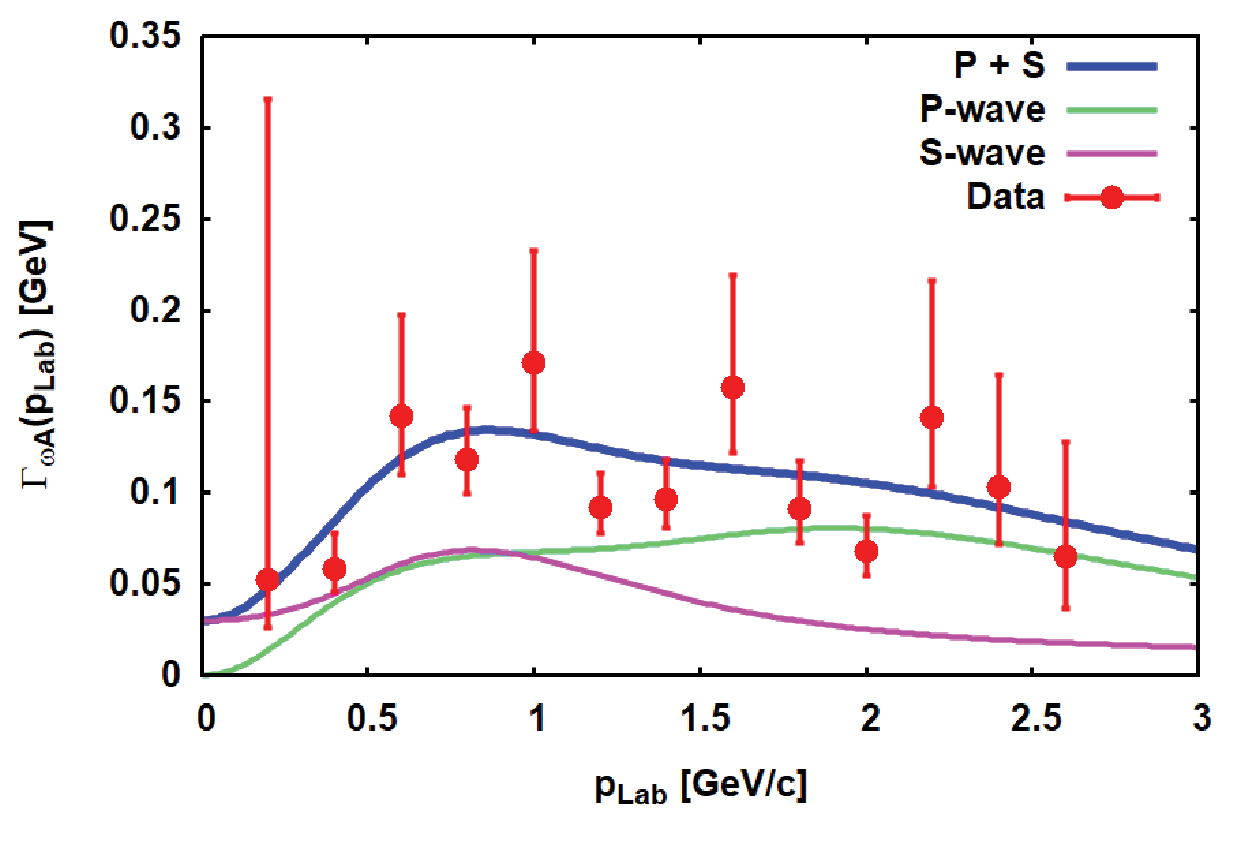}
\includegraphics[width=7cm,clip]{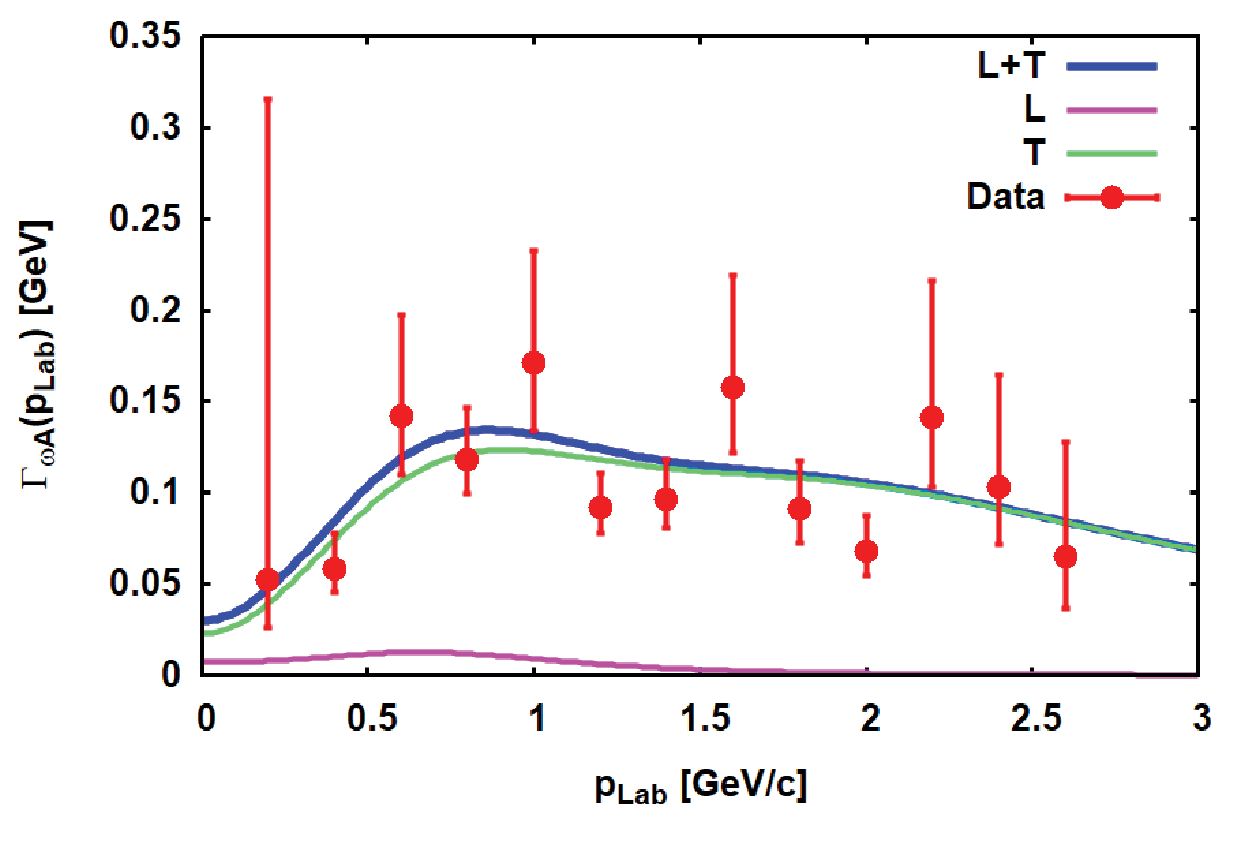}
\includegraphics[width=7cm,clip]{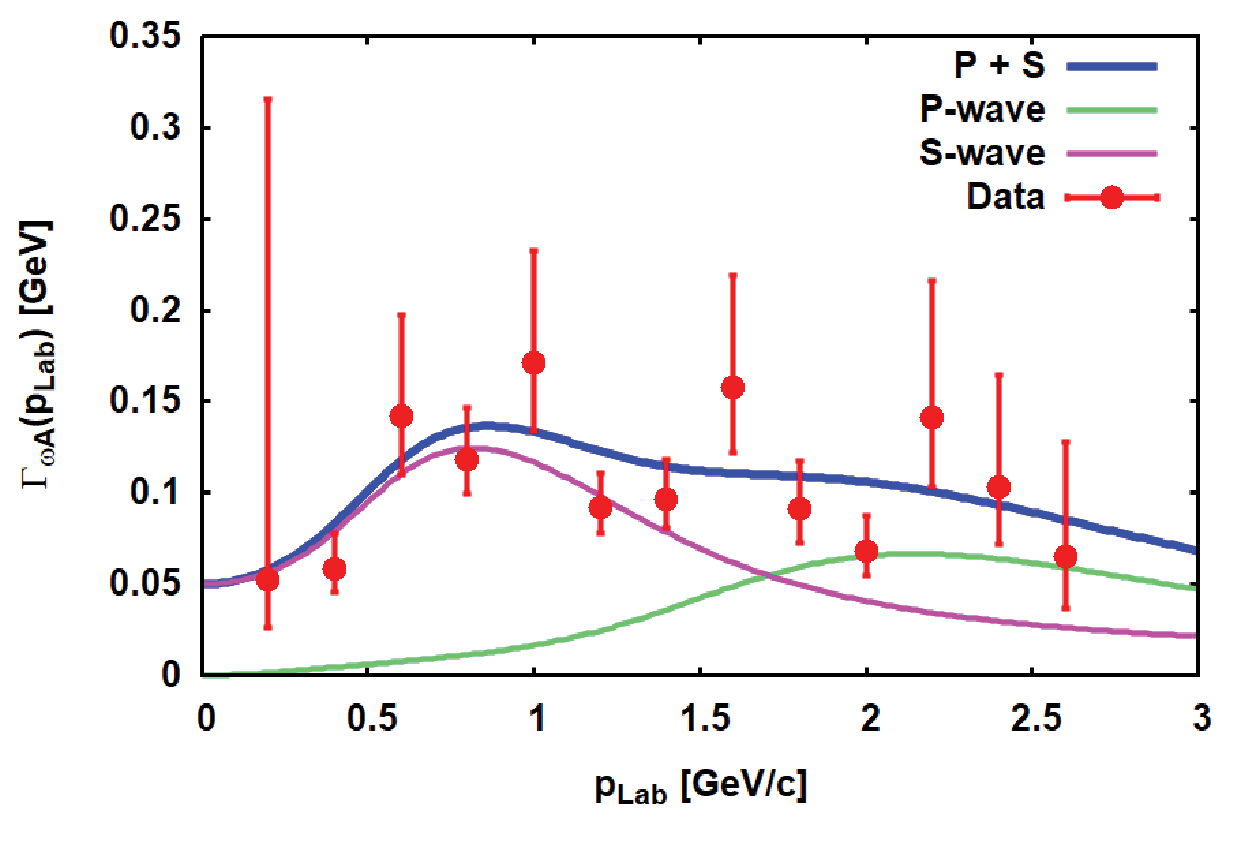}
\includegraphics[width=7cm,clip]{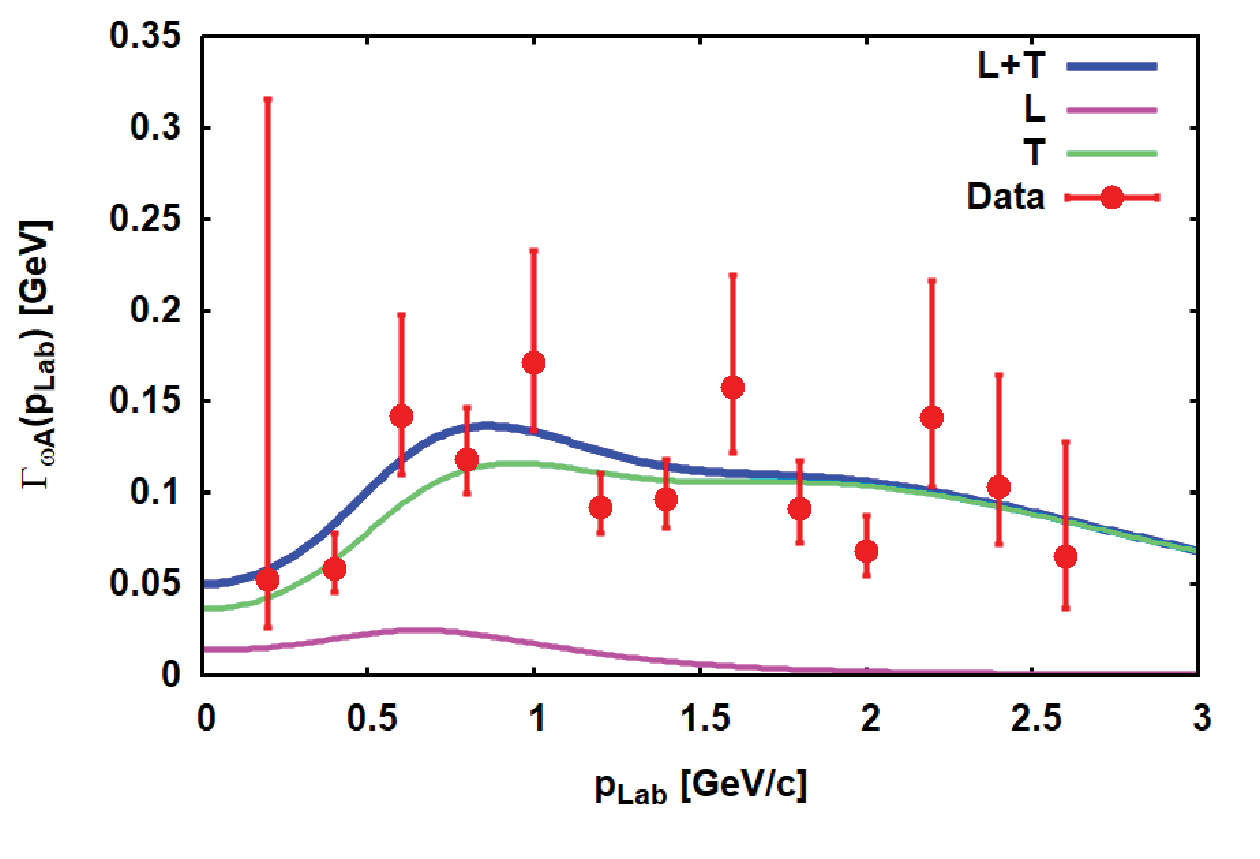}
\includegraphics[width=7cm,clip]{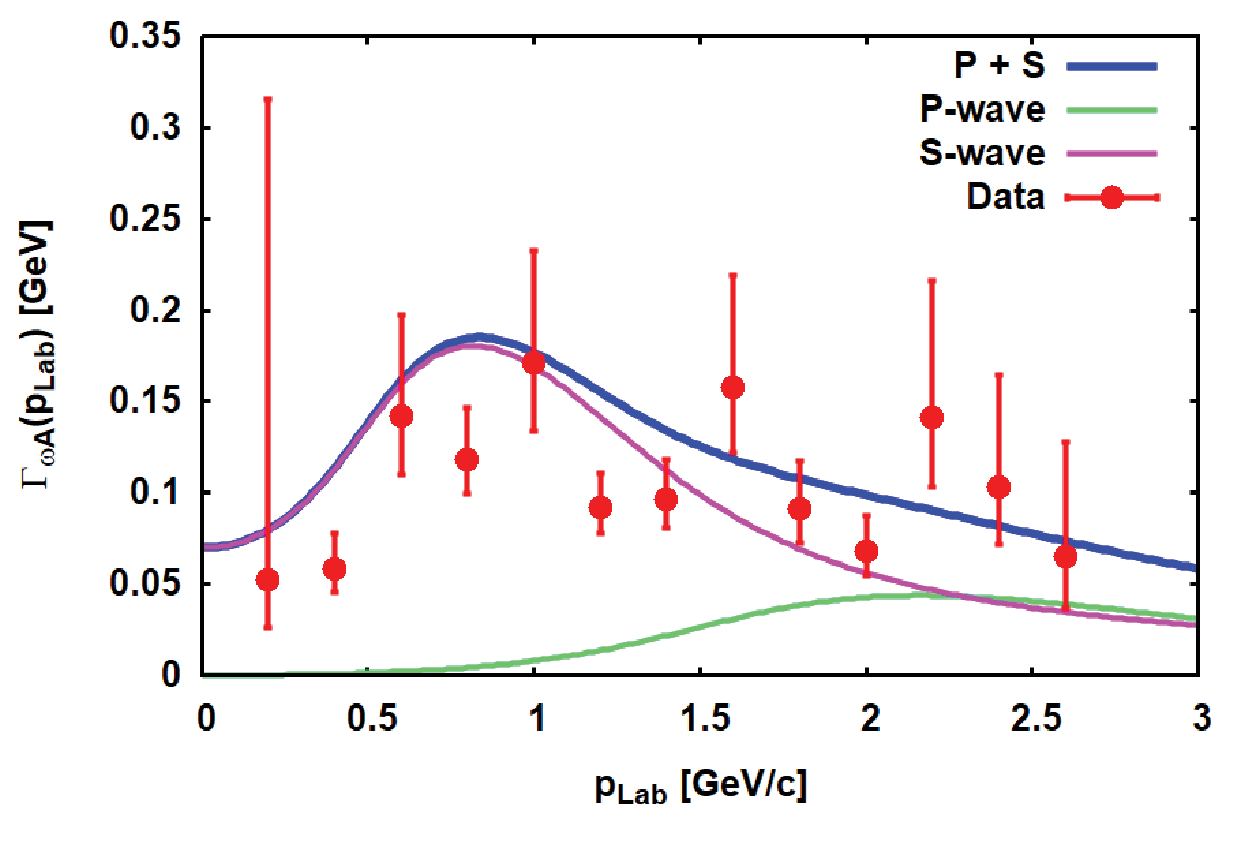}
\includegraphics[width=7cm,clip]{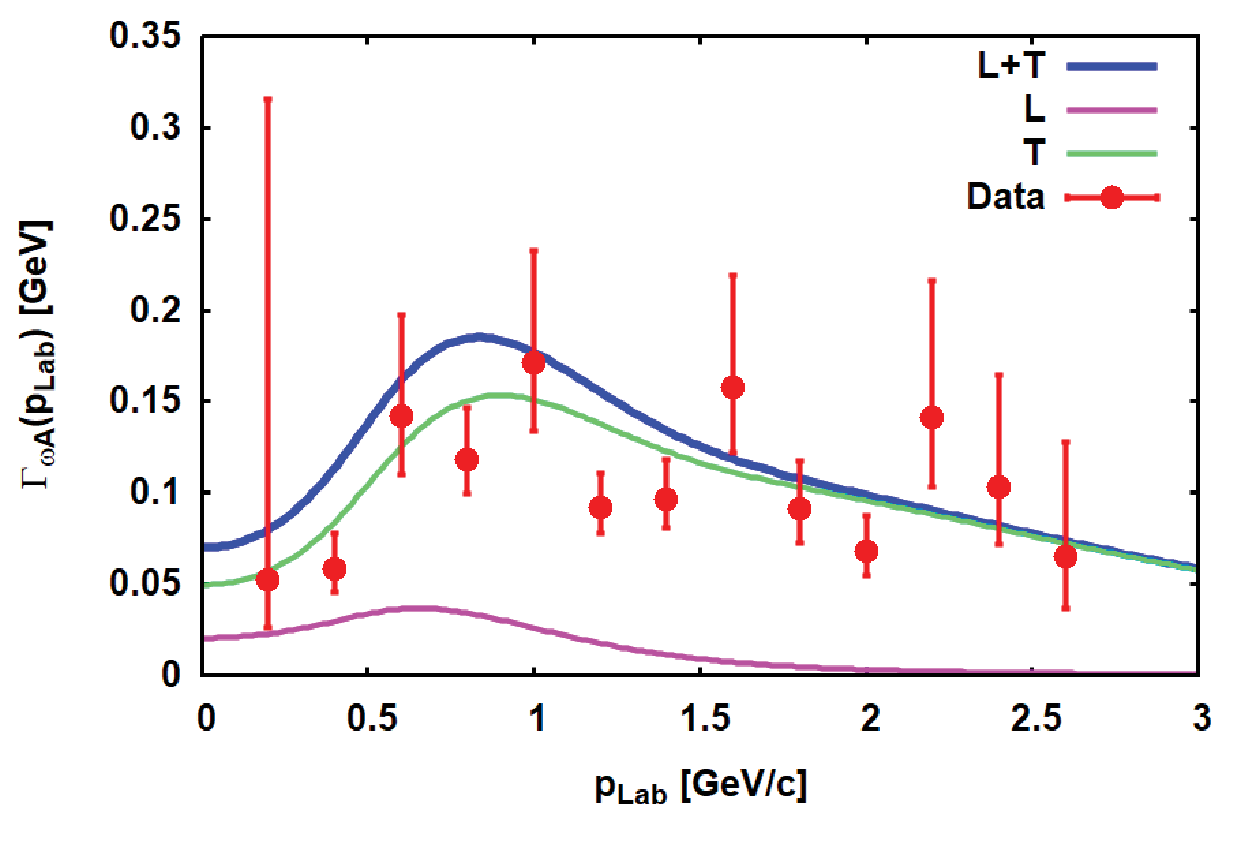}
\caption{Decomposition of the $\omega +{}^{93}$Nb widths into P--wave (P) and S--wave (S) modes (left) and longitudinal (L) and transversal (T) components (right) at central density $\rho_A=0.140fm^{-3}$ of $^{93}$Nb. Results of calculations are shown for $\Gamma_{thr}=30$~MeV (upper row),
$\Gamma_{thr}=50$~MeV (center row), and $\Gamma_{thr}=70$~MeV (bottom row). Data are taken again from Ref. \cite{Friedrich:2016cms}. }
\label{fig:GwPScLT}
\end{figure}

\subsection{Dependence of Longitudinal and Transversal Self--Energies on the Threshold Value}
As emphasized repeatedly before, the value of the width at threshold is decisive for the content of S--wave self--energies and, as such, for the composition of the omega in--medium self--energy tensors as a whole.
Therefore, it is of high interest to investigate the dependence of the theoretical distribution on the threshold value, thus further encircling systematic uncertainties in the modelling. For that purpose, the constrained scenario (CS) is used by performing a series of calculations with varying $\Gamma_{\omega A}(0)=\Gamma_{thr}$ from $\Gamma_{thr}=30$~MeV to $\Gamma_{thr}=70$~MeV.

\begin{figure}[h]
\centering
%\sidecaption\overrightarrow{}
\includegraphics[width=7cm,clip]{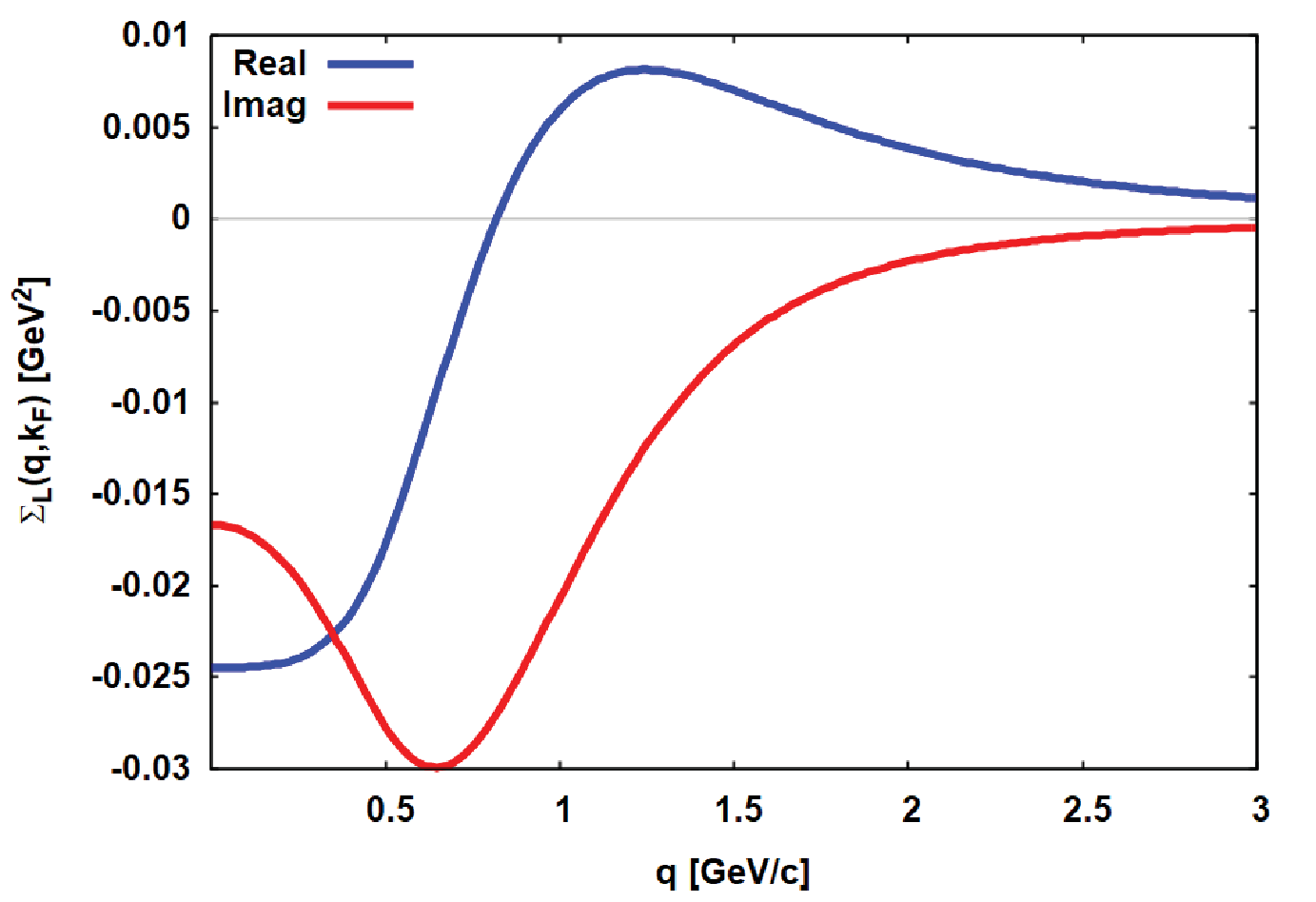}
\includegraphics[width=7cm,clip]{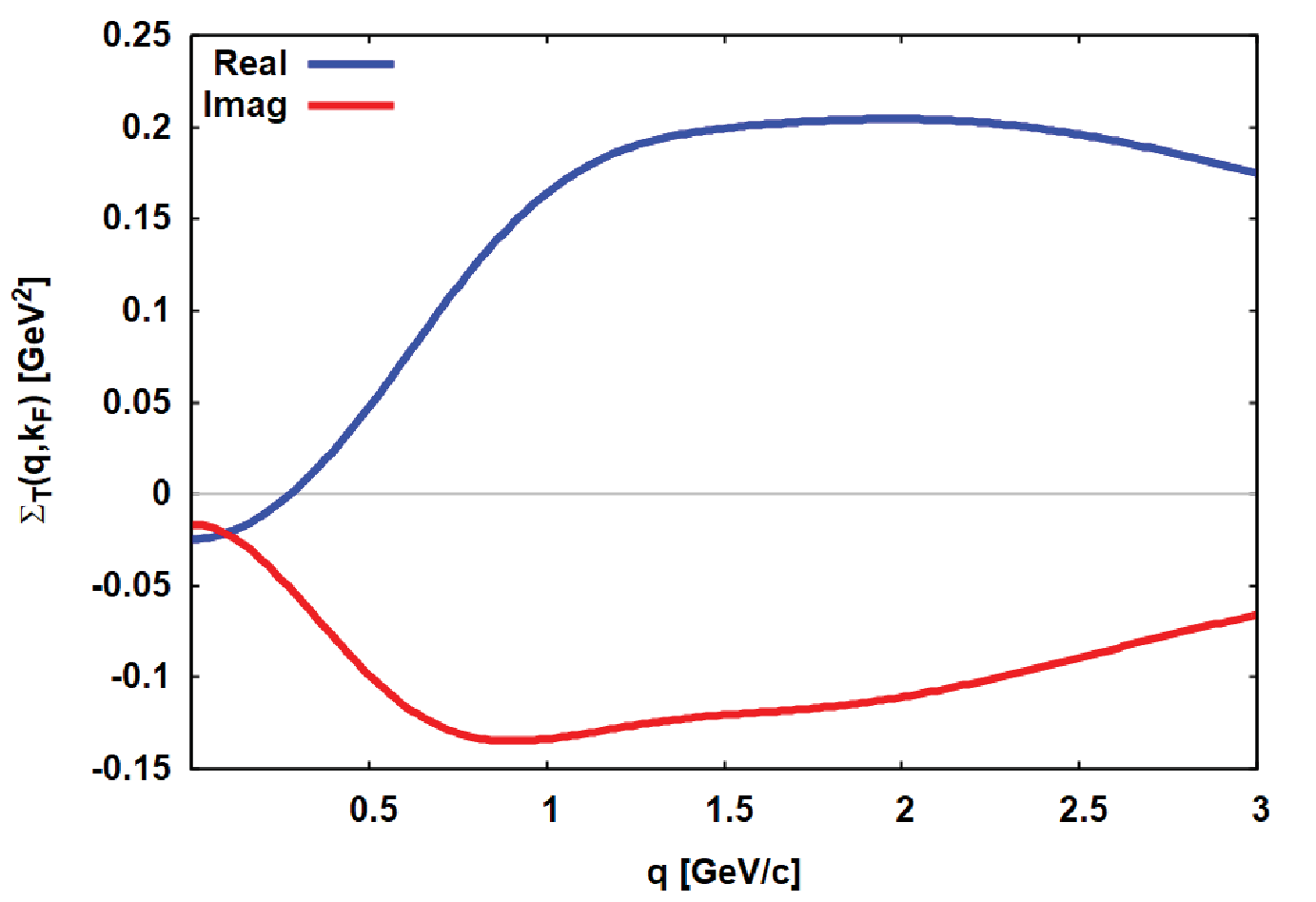}
\includegraphics[width=7cm,clip]{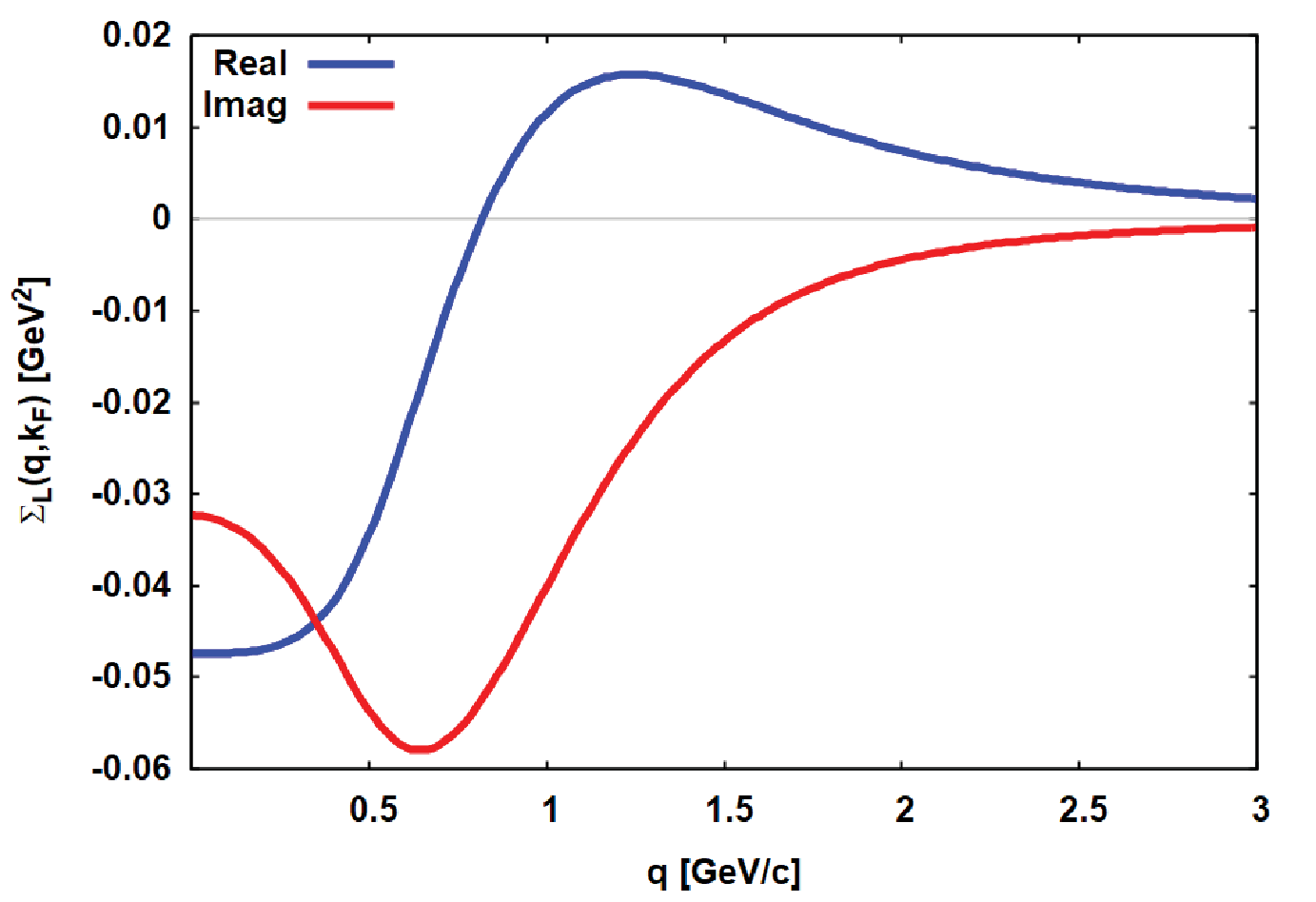}
\includegraphics[width=7cm,clip]{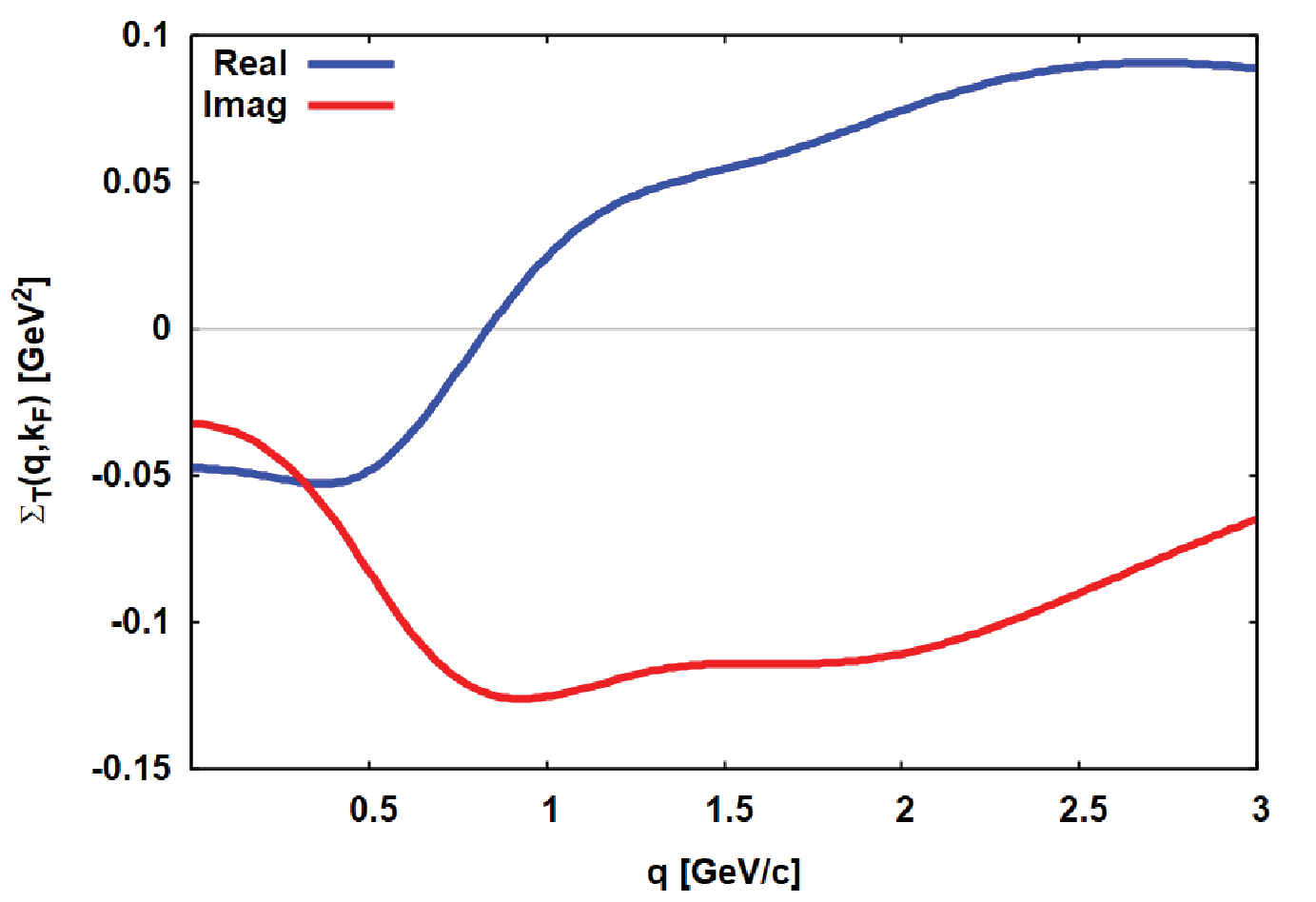}
\includegraphics[width=7cm,clip]{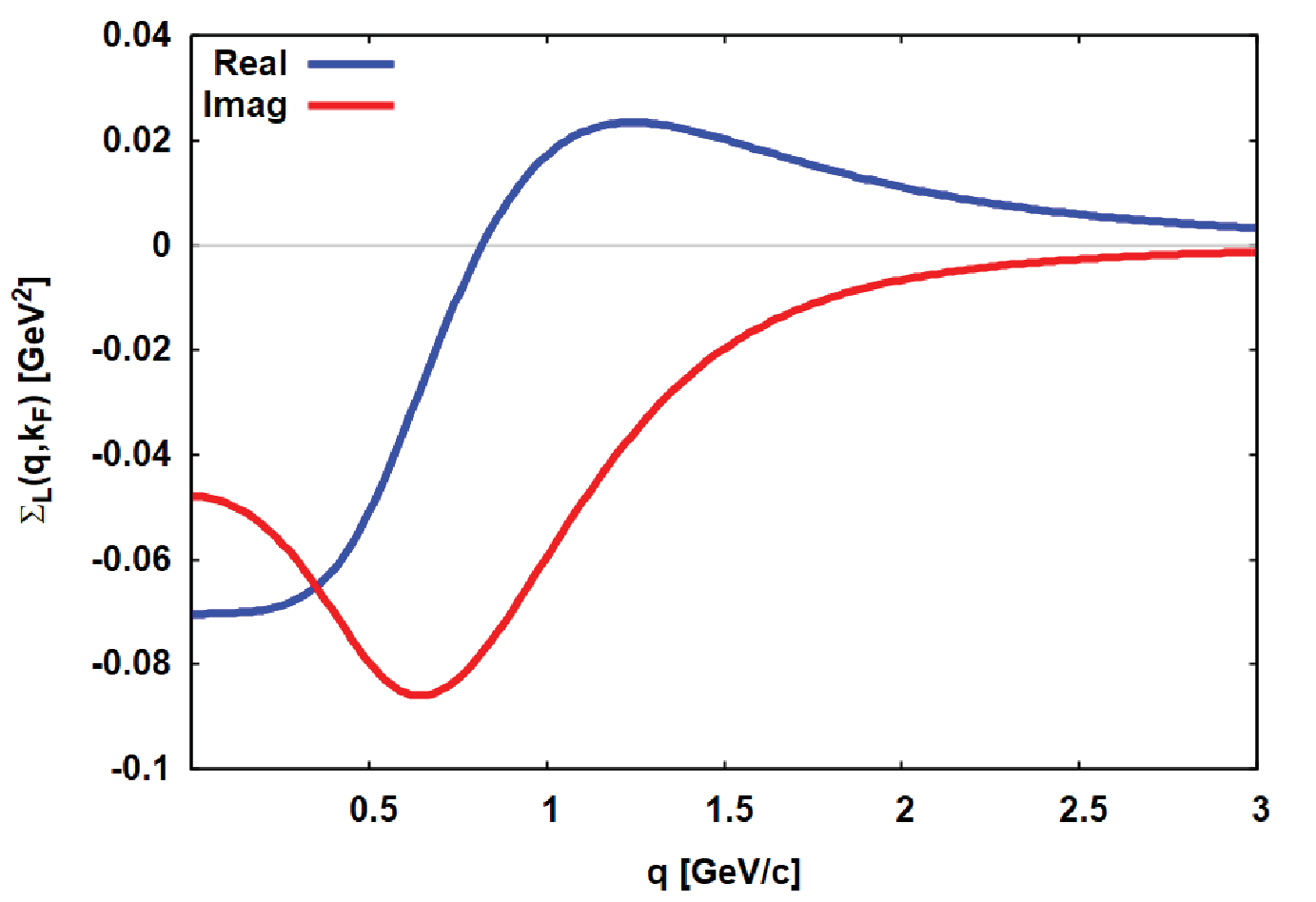}
\includegraphics[width=7cm,clip]{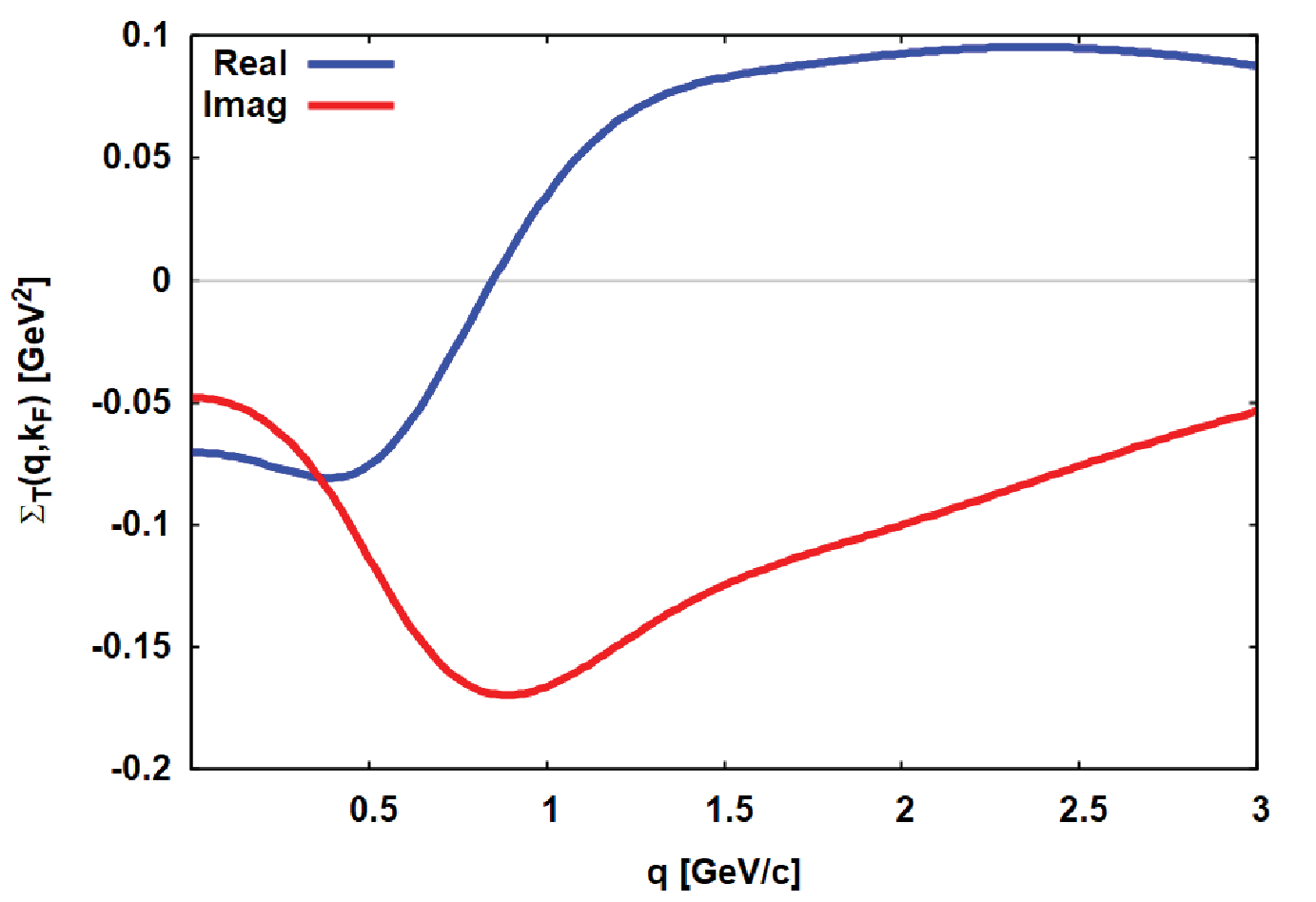}
\caption{Dependence of self--energies on $\Gamma_{thr}$. In the left column the real and imaginary parts of longitudinal (L) and transversal (T) self--energies are shown for the arbitrarily chosen threshold widths $\Gamma_{thr}=30$~MeV (top row),
$\Gamma_{thr}=50$~MeV (center row), and $\Gamma_{thr}=70$~MeV (bottom row). All results
were obtained for the central density $\rho_A=0.140fm^{-3}$ of $^{93}$Nb.  }
\label{fig:SwCSLT}
\end{figure}

\begin{table}
  \centering
\begin{tabular}{|c|c|c|c|c|c|c|c|}
  \hline
  $\Gamma_{\omega A}(0)$/MeV &$\Gamma^{(A)}_{0}$/MeV & $\mathcal{P}_S$ & $\mathcal{P}_P$ &$\mathcal{P}_L$& $\mathcal{P}_T$& $\chi^2/ndf$ & Type \\\hline
    48.41 & 39.73 & 0.576 & 0.424 & 0.096 & 0.904 & 1.014& US\\
    30.00 & 21.32 & 0.313 & 0.687 & 0.052 & 0.948 & 1.045& CS\\
    40.00 & 31.32 & 0.448 & 0.542 & 0.076 & 0.924 & 1.045& CS\\
    50.00 & 41.32 & 0.600 & 0.400 & 0.100 & 0.900 & 1.008& CS\\
    60.00 & 51.32 & 0.701 & 0.299 & 0.117 & 0.883 & 1.045& CS\\
    70.00 & 61.32 & 0.781 & 0.219 & 0.130 & 0.870 & 2.726& CS\\
  \hline
\end{tabular}
  \caption{Total in--medium $\omega +{}^{93}$Nb width (first column), the contained many--body part (second column) at the nuclear center are shown together with the fractional S--wave ($\mathcal{P}_S$, third column) and P--wave ($\mathcal{P}_P$, fourth column) and the fractional longitudinal ($\mathcal{P}_L$, fifth column) and transversal ($\mathcal{P}_T$, sixth column) self--energy contributions.  The first line contains the results obtained from the unconstrained fit, the other lines contain the results for the constrained fits, where the width at threshold was fixed artificially to the listed values. In all cases contributions of S--wave and P--wave resonance were included. The achieved $\chi^2$ per number of degrees of freedom (ndf) are displayed in the 6th column. The last column indicates the type of fitting procedure, US for unconstrained scenario and CS for constrained scenario.   }\label{tab:Tab2}
\end{table}

In Fig.\ref{fig:GwPScLT}, results of CS calculations are displayed for $\Gamma_{thr}=30$~MeV, $\Gamma_{thr}=50$~MeV, and $\Gamma_{thr}=70$~MeV, respectively. In Tab. \ref{tab:Tab2} the dependence of the fractional S--wave and P--wave contributions on $\Gamma_{thr}$ are displayed. Both Fig.\ref{fig:GwPScLT} and Tab. \ref{tab:Tab2} clearly demonstrate the pronounced dependence of S--wave and P--wave contents on $\Gamma_{thr}$, also in comparison to the US--results, Fig.\ref{fig:GwPSuLT}.

The delicate balance of S-- and P--wave components is emphasized when varying the assumed threshold value $\Gamma_{thr}$. Comparing in Fig. \ref{fig:GwPScLT} the results for the three pre--chosen values of the threshold width it is seen in the left column that with increasing $\Gamma_{thr}$ the S--wave parts gain increasingly more weight. The reason is obvious because the width at threshold is fully determined by the S--wave components only. However, strengthen the S--wave contributions implies to weaken the P--wave parts. Finally, beyond $\Gamma_{thr}\sim 60$~MeV that trade--off leads to an underestimation of the width at $p_{lab}>1.5$~GeV/c. In Tab. \ref{tab:Tab2} this effect is reflected by the sudden jump to $\chi^2/ndf = 2.726$. From Tab. \ref{tab:Tab2} and comparing Fig. \ref{fig:GwPSuLT} and Fig. \ref{fig:GwPScLT} it is also seen that $\Gamma_{thr}=48.41$~MeV obtained by the unconstrained scenario complies very well with the results found in the constrained fit with $\Gamma_{thr}=50$~MeV.

The real parts of the longitudinal and transversal self--energies, shown in Fig.\ref{fig:SwCSLT}, are seen to vary considerably more pronounced under variations of the threshold width than the imaginary parts. A common global feature of the longitudinal and transversal components is a change from attraction at low energies to repulsion at higher energy. Overall, however, the transversal real components are of a much larger magnitude than the longitudinal ones. Moreover, the real transversal self--energies are affected remarkably not only at low energies but over the whole energy range. The explanation of that effect is that increasing the threshold width gives a stronger weight to the S--wave self--energies, which is compensated by decreasing the P--wave parts. Since the latter account for a large part of the transversal self--energies (see Eq. \eqref{eq:SigmaT}), they are reacting stronger on variations of the P--wave components.

\subsection{Spectral Distributions in Infinite Nuclear Matter}\label{ssec:SpecD}

An issue discussed heavily and controversially for a long time in the literature is the dependence of the spectral properties of mesons and especially vector mesons in nuclear matter and whether and how they are observable experimentally, see e.g. \cite{Peters:1997va,Muehlich:2006nn,Rapp:2009yu,Metag:2017yuh}. As discussed in App. \ref{app:freeVM}, the separation of vector meson in--medium self--energies into longitudinal and transversal components induces a corresponding separation of the propagator into a longitudinal and a transversal propagator. Consequently, in matter a vector meson acquires two separate spectral functions with their own kind of dependencies on energy $w$, momentum $q$, and nuclear density $\rho$. The longitudinal and transversal spectral functions of the omega--meson are defined by
\begin{eqnarray}
\label{eq:fullspec}
A_{L/T}(w,\mathbf{q}|\rho) = - \frac{1}{\pi} \;{\rm Im}\left(
        \frac{1}{w^2-\mathbf{q}^2 - m^2_{\omega} -\Sigma_{L/T}(w,\mathbf{q},k_F)
        -\Sigma_{free}(w,\mathbf{q})}
\right) \quad .
\end{eqnarray}
All medium--dependent effects are contained in $\Sigma_{L/T}(w,\mathbf{q},k_F)$. Hence, since by definition and construction $\Sigma_{L/T}=0$ at $\rho=0$, both distributions converge to the free--space distribution
\be
A_{free}(w,\mathbf{q})=- \frac{1}{\pi} \;{\rm Im}\left(
        \frac{1}{w^2-\mathbf{q}^2 - m^2_{\omega}
        -\Sigma_{free}(w,\mathbf{q})}\right)\quad ,
\ee
with the self--energy in free space $\Sigma_{free}=-im_\omega \Gamma_{free}$ where the real part is supposed to be absorbed into the physical mass $m_\omega$.

\begin{figure}[h]
\centering
%\sidecaption\overrightarrow{}
\includegraphics[width=7cm,clip]{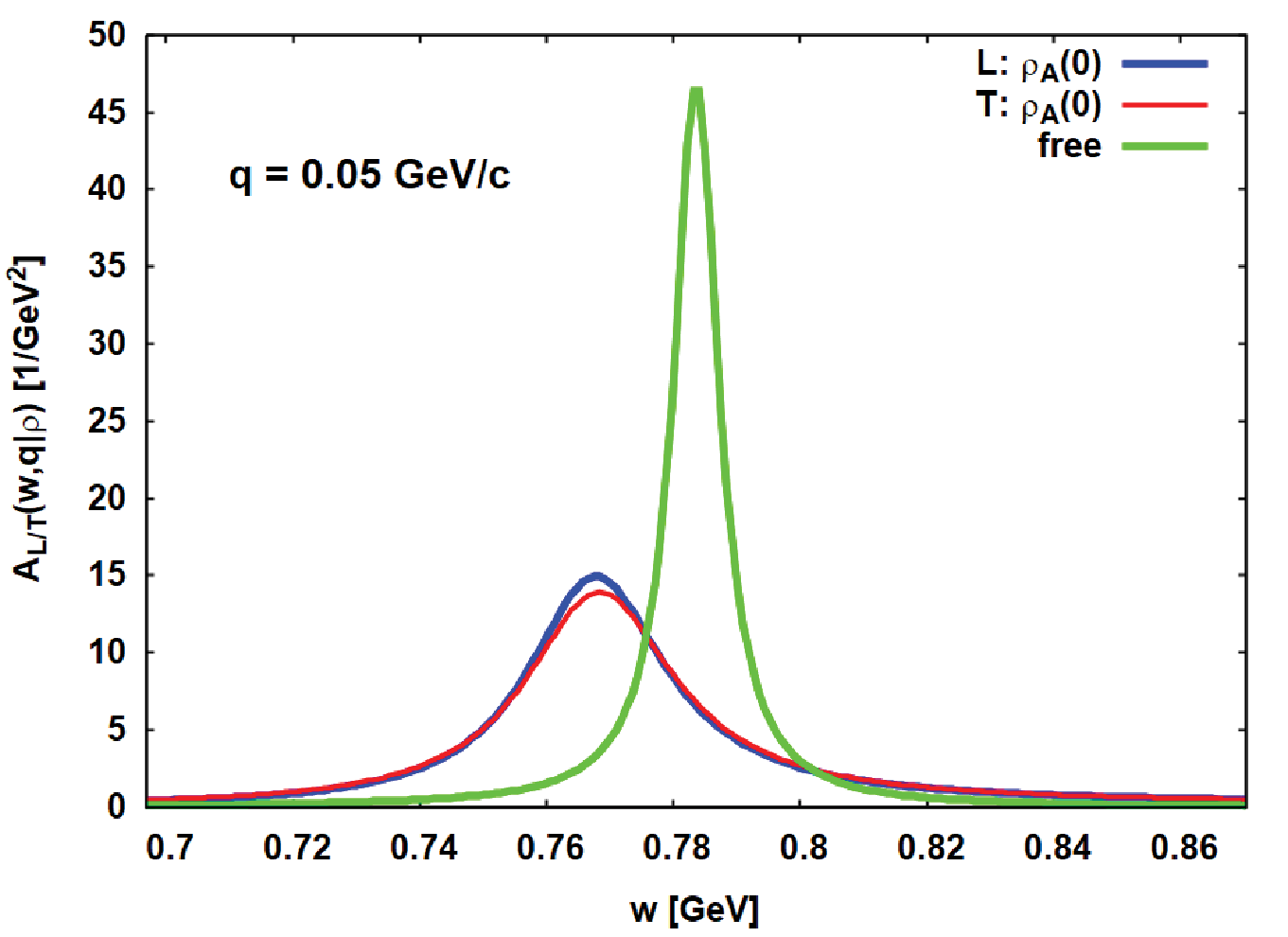}
\includegraphics[width=7cm,clip]{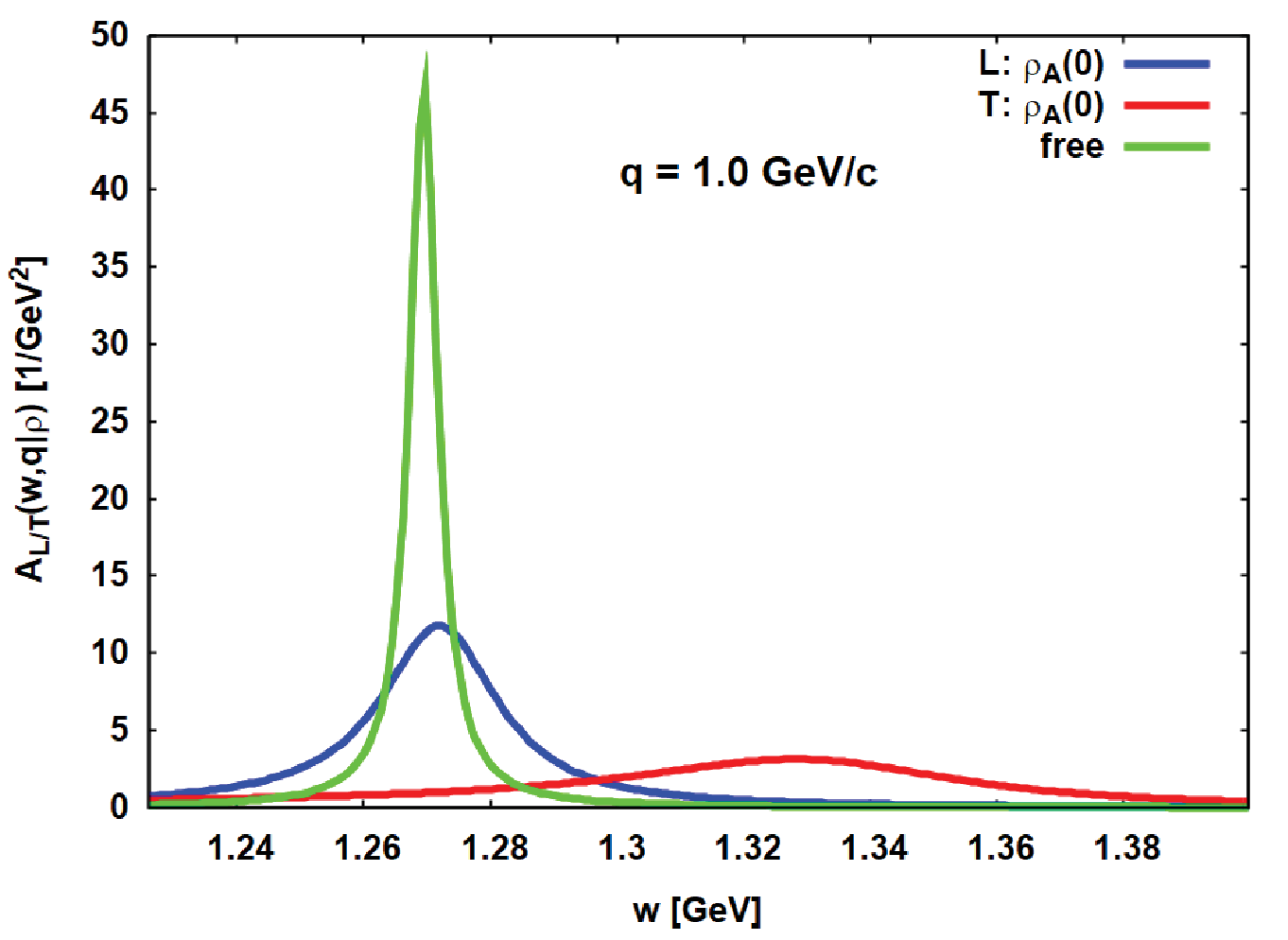}
\includegraphics[width=7cm,clip]{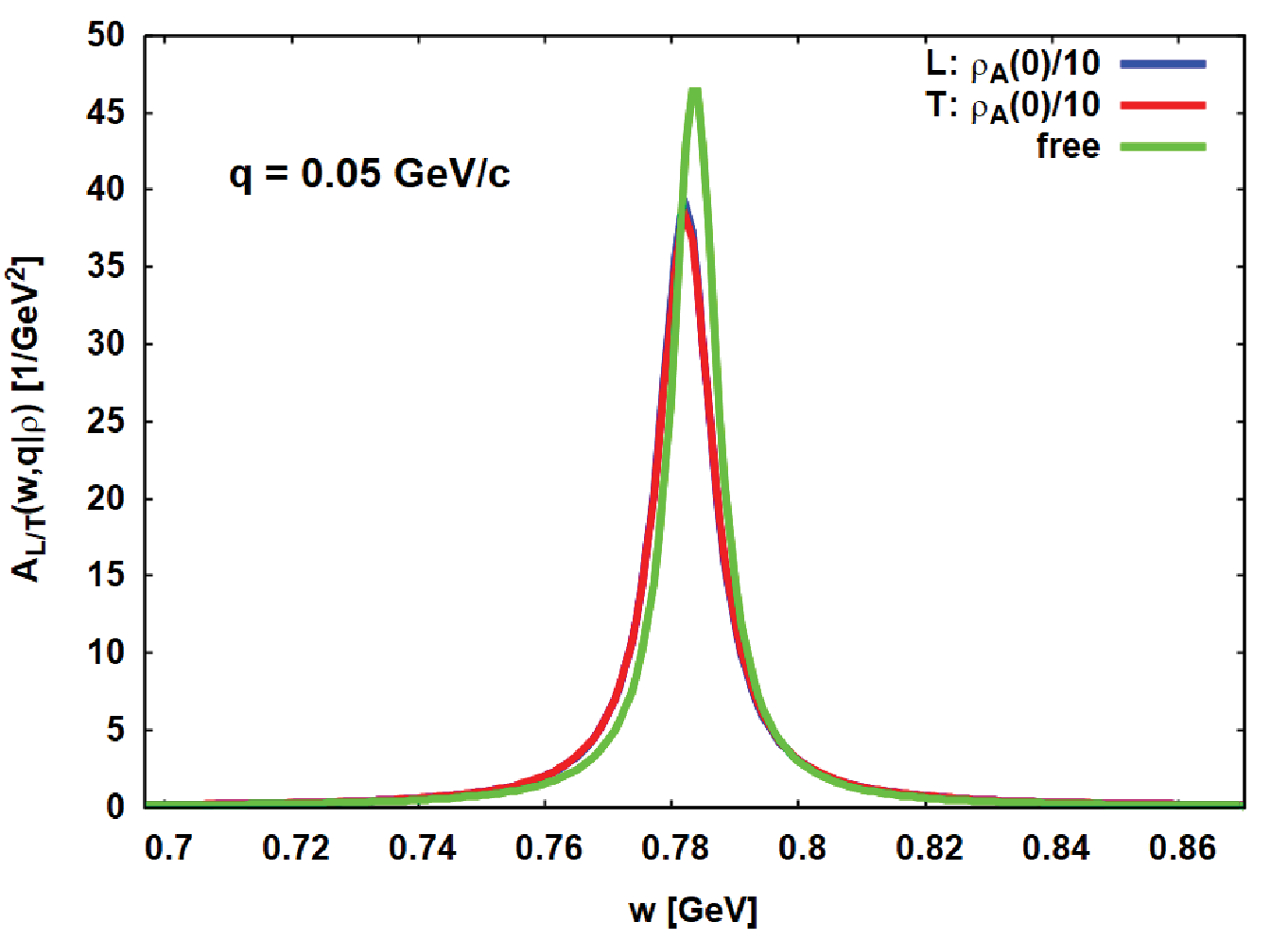}
\includegraphics[width=7cm,clip]{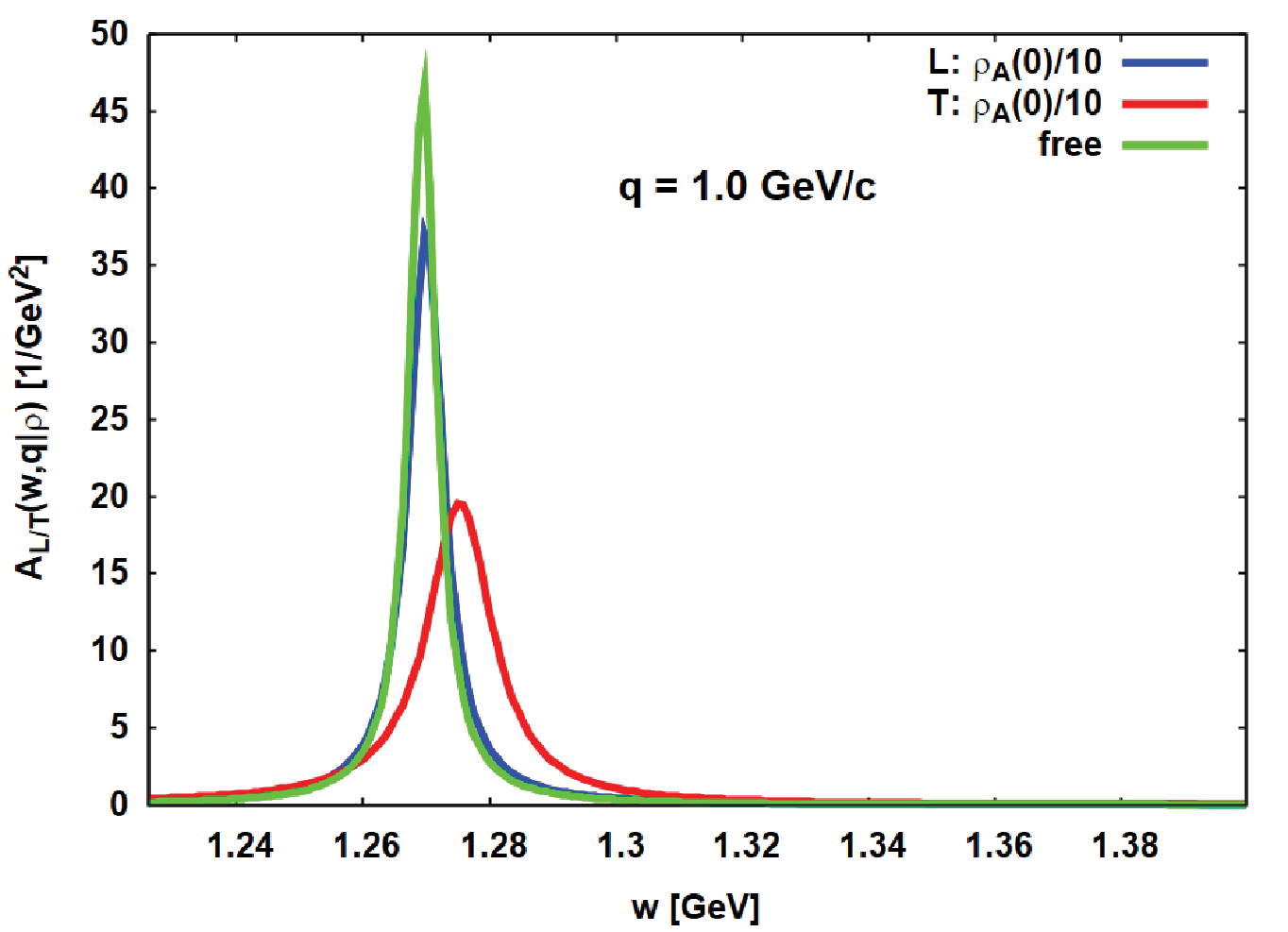}
\caption{Spectral distributions of an omega--meson in nuclear matter as functions of the energy $w=\sqrt{s}-\sqrt{s_{thr}}$ at fixed momentum $q=0.05$~GeV/c (left) and $q=1$~GeV/c (right) and at the central density $\rho_A=0.140fm^{-3}$ of $^{93}$Nb (upper row) and at $\rho=\frac{1}{10}\rho_A$ (lower row). For comparison, the spectral distribution in free space is also displayed.  }
\label{fig:SpecD}
\end{figure}

The spectral distributions are functions of the 4--momentum $p=\left(w,\mathbf{q} \right)$. In Fig.\ref{fig:SpecD} distributions are displayed as functions of the energy $w$ for two representative values of the 3--momentum $q=0.05$~GeV/c and $q=1$~GeV/c, respectively, both at central density of $^{93}$Nb $\rho_A=0.140 fm^{-3}$ and at low density $\rho=\frac{1}{10}\rho_A=0.014 fm^{-3}$, representative for the nuclear surface region. The in--medium distributions are compared to the spectral function in free space.

First of all, irrespective of the momentum the distributions approach the free--space spectral function with decreasing density, as seen by comparing the upper and lower rows of Fig.\ref{fig:SpecD}. Thus, at zero density the theoretical distributions merge to the expected common limiting case.

Comparing the left and right column of Fig.\ref{fig:SpecD}, one finds a pronounced dependence on the momentum. For $q=0.05$~GeV/c and central density (upper left) both the longitudinal and the transversal distribution are moved to energies to the left of the free--space spectral function, i.e. the in--medium distributions are shifted to lower energies. At high momentum, $q=1$~GeV/c (upper right), the longitudinal and the transversal distributions behave quite differently: While $A_L$ is centered in the same energy region as $A_{free}$, the transversal distribution is shifted to higher energies with a width $\Gamma_T\gg \Gamma_L$. In the low density region (lower row), these differences regarding the momentum dependence prevail, although on a less pronounced level. For $q=0.05$~GeV/c the two in--medium distributions are overlapping closely with $A_{free}$ but with a slightly larger width. At $q=1$~GeV/c, $A_L$ has almost approached the free--space distribution while $A_T$ is still clearly separated with a broader width.

Thus, the answer to the question whether there is an in--medium mass shift to lower masses (often discussed as a signal of approaching the chiral limit) or an upward shift to larger masses is ambiguous because it depends obviously on the conditions under with the system is probed. The model calculations predict that "mass shifts" will occur in both directions, depending on the momentum brought into the system. The reason for that -- on first sight conflicting -- behaviour is found in the real parts of the polarization self--energies. A look to Fig.\ref{fig:SwCSLT} reveals immediately that over the energy--momentum range the longitudinal and transversal real self--energies change character from attractive to repulsive. For off--shell conditions the change of sign in the real L/T self--energies will occur in general in different energy--momentum regions and with different magnitude, thus producing apparent "mass shifts" in a seemingly conflicting manner.

\section{Dispersive $\omega$ Self-Energies in a Finite Nucleus}\label{sec:FiniteNucleus}

\subsection{Local Density Approximation}
Infinite matter calculations result in self--energies depending parameterically on the nucleon number density $\rho_N=\{\rho_p,\rho_n\}$ and by the well known relation $\rho_{p,n}\sim k^3_{F_{p,n}}$ on the Fermi momentum $k_F=\{k_{F_p},k_{F_n} \}$. As in \cite{Rodriguez-Sanchez:2020aht} the local density approximation (LDA) is used for the mapping to a finite nucleus by replacing the densities by the radial nuclear ground state density distributions of the nucleus under consideration, $\rho_N\mapsto \rho_A(\mathbf{r})$ and accordingly $k_{F}\mapsto k_{F_A}(\mathbf{r})$. In this way, the original density dependence is translated into a radial dependence, imprinting the nuclear density profile on the self--energies.

\begin{figure}[h]
\centering
%\sidecaption
\includegraphics[width=6cm,clip]{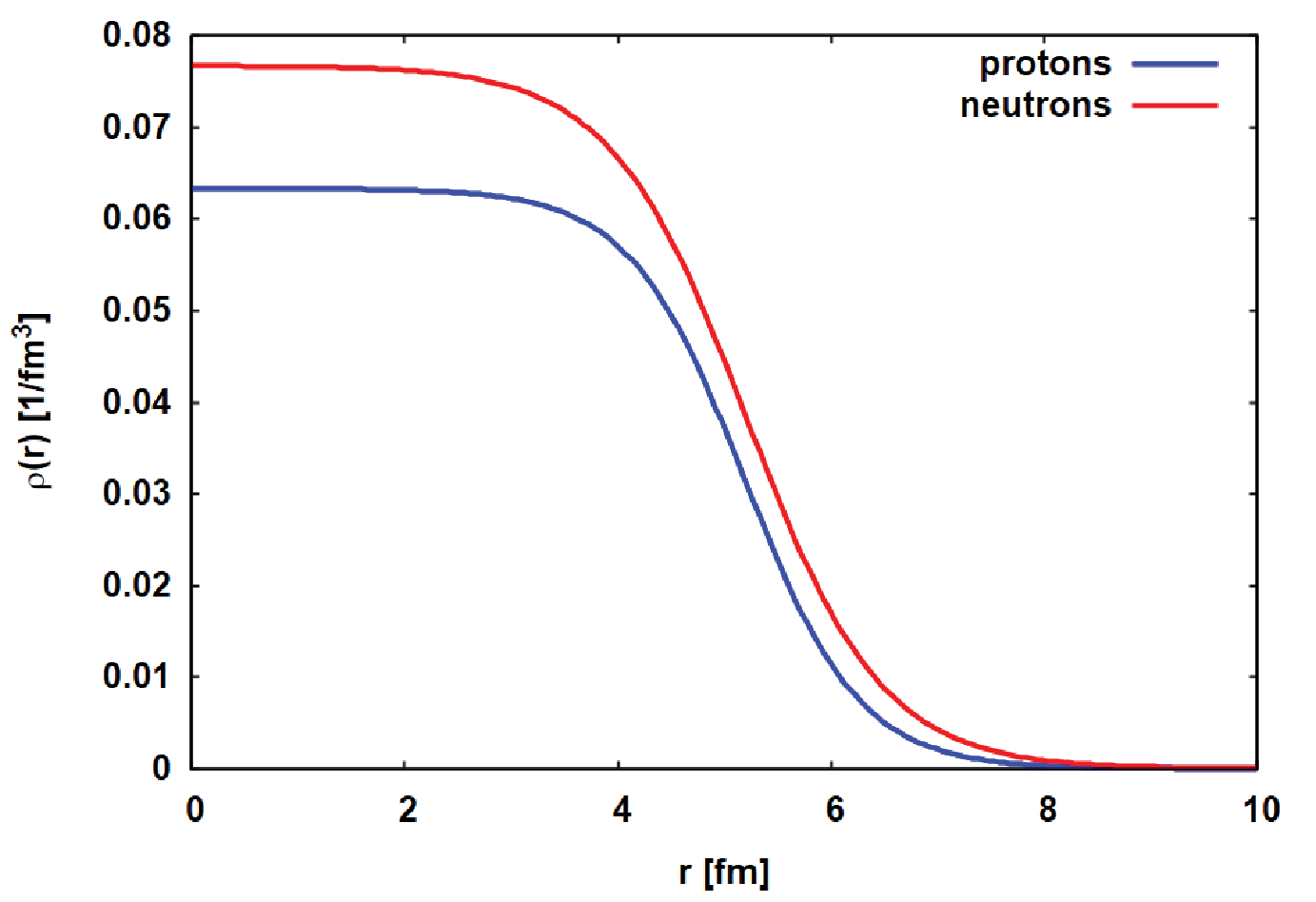}
\includegraphics[width=6cm,clip]{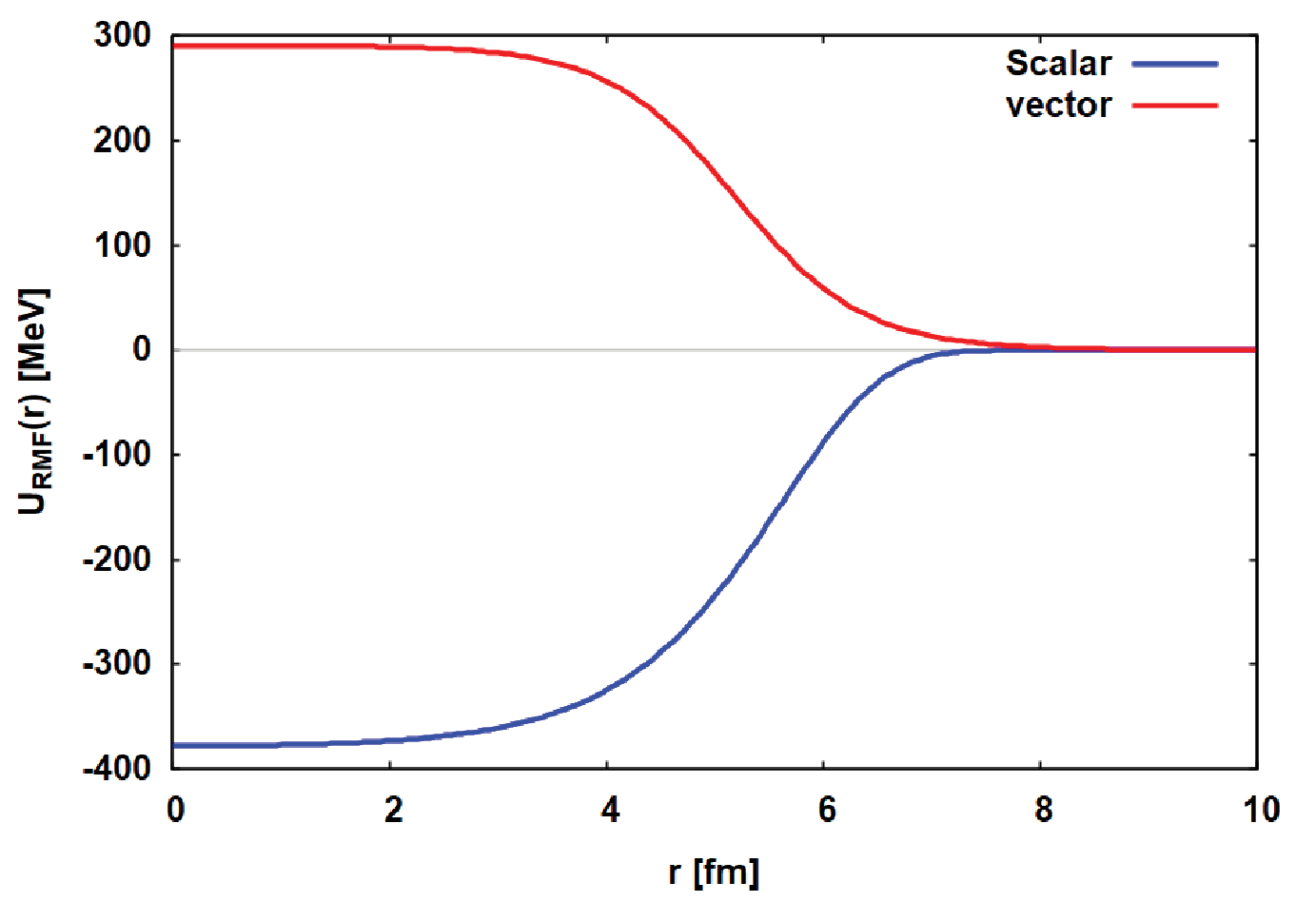}
\caption{RMF--Hartree proton and neutron ground state densities (left) of $^{93}$Nb and related scalar and vector Hartree mean--fields (right) entering in LDA into the self--energies.  }
\label{fig:gsdNB}
\end{figure}

The use of the LDA with densities and potentials from self--consistent covariant Hartree calculations is an essential and important difference to previous investigations. Isoscalar and
isovector scalar and vector mean--fields are important prerequisites for a realistic treatment of charge--asymmetric nuclei like
$^{93}$Nb. The proton and neutron RMF--Hartree ground state densities of $^{93}$Nb are displayed in Fig.\ref{fig:gsdNB} together with the (isoscalar) scalar and vector Hartree mean--field.

\subsection{Self--Energies for $\omega + {}^{93}Nb$}
\begin{figure}[h]
\centering
%\sidecaption\overrightarrow{}
\includegraphics[width=15cm,clip]{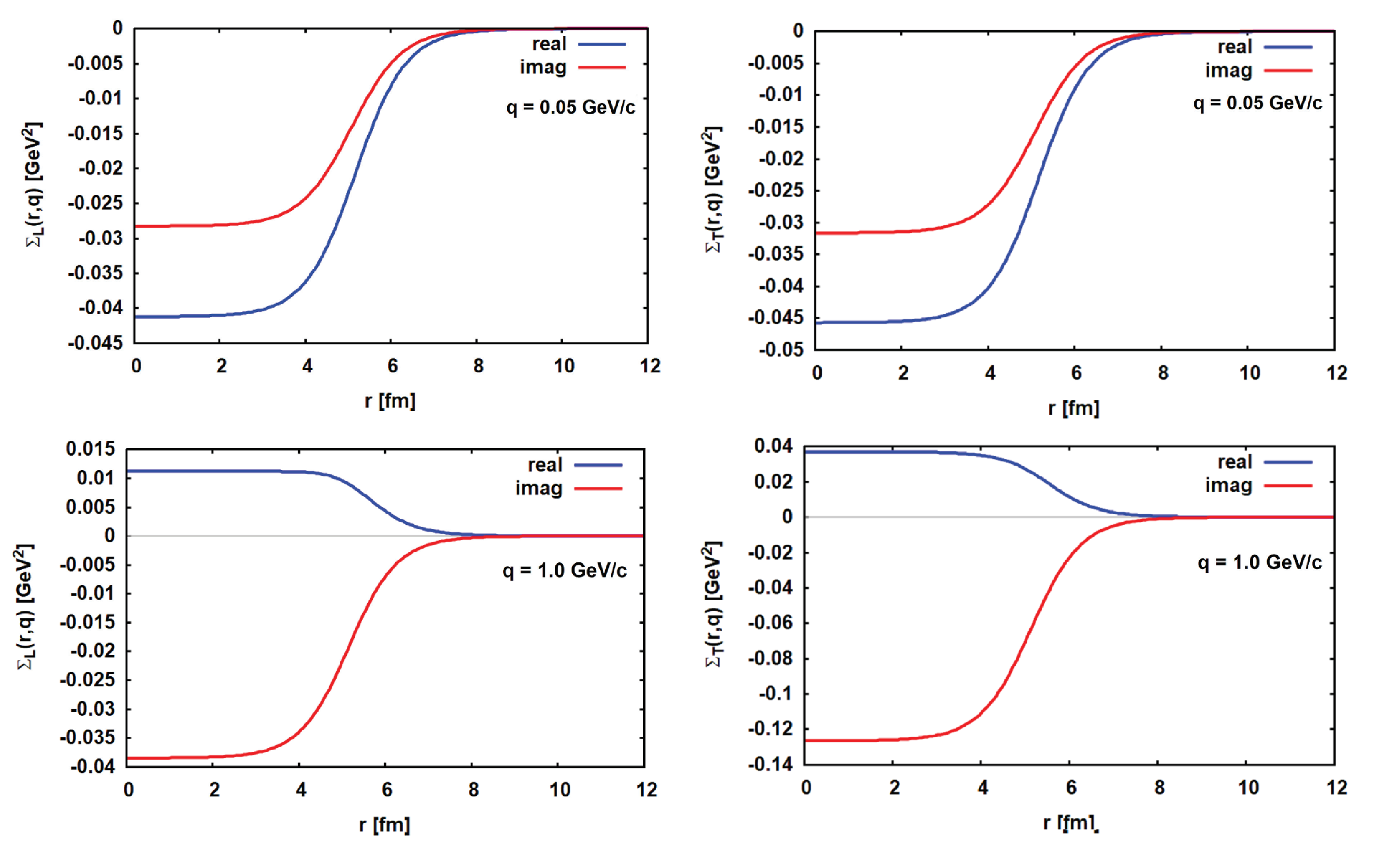}
\caption{Radial longitudinal (left column) and transversal (right column) self--energies in $^{93}$Nb. LDA--results are displayed for
$q=0.05$~GeV/c (upper row) and $q=1$~GeV/c (lower row), respectively, illustrating the change from attraction to repulsion with increasing energy.}
\label{fig:SrLT}
\end{figure}

In this section, longitudinal and transversal self--energies in $^{93}$Nb are considered in LDA using the densities of Fig.\ref{fig:gsdNB}. In Fig.\ref{fig:SrLT} the self--energies are displayed as functions of the radius for $q=0.05$~GeV/c and $q=1$~GeV/c, corresponding to kinetic energies $T_{cm}\sim 1.60$~MeV and $T_{cm}\sim 490$~MeV, respectively. In the low momentum/energy case, the P--wave self--energies are still negligible. Hence, the longitudinal and transversal self--energies are of similar magnitude, as evident by comparing the plots in the upper row. The low--energy real self--energies are attractive.
A completely different situation is encountered at $q=1$~GeV/c: Now the P--wave components are sizable, modifying the transversal self--energies considerably. As a result, the longitudinal and transversal self--energies are of a quite different magnitude. Moreover, the real self--energies have changed sign to repulsion at this higher energy.

By visual inspection, one may conclude that the shapes follow closely the Niobium ground states densities, Fig.\ref{fig:gsdNB}, thus seemingly confirming the afore mentioned density--scaling laws for the width, used in various studies. However, a closer look reveals a more complex relation. Close to threshold, the radial form factors of the real self--energies resemble indeed closely the underlying $^{93}$Nb RMF--ground state density but start to deviate rapidly with increasing energy. The radial from factors of the imaginary self--energies, however, differ at all energies from the Niobium ground state density, questioning simple density--scaling laws.

A pronounced behaviour is encountered around the momenta where the real self--energies change sign: The transition from attraction to repulsion is governed by interfering S--wave and P--wave contributions differing in sign and magnitude which at a certain critical momentum results in a vanishing real part before the repulsive domain is entered. For example, in a narrow window around $q=0.8$~GeV/c the real longitudinal self--energy changes sign by going though a mixed phase of a superposition of an attractive volume part, whose magnitude gradually decreases, and a repulsive component of complementary increasing strength with a form factor changing with energy from a surface--peaked to a volume shape.
At  $q\simeq 0.815$~GeV/c the transition point is reached, the real longitudinal self--energy vanishes, and afterwards builds up a repulsive volume shape. At $q=1$~GeV/c, in fact, a trace of this transitional behaviour is still visible.
A similar scenario is observed for the real transversal self-energy, having passed this transient region already at a lower momentum, $q\sim 0.7$~GeV/c.

In the asymptotic region beyond the nuclear radius the deviations between self--energy and density form factors steadily grow. The reason is that in the tail region the differences between the asymptotic shapes of proton and neutron densities become visible, the latter being determined essentially by the differences in chemical potentials.

\subsection{Schr\"{o}dinger-like $\omega$ Potentials and Low--energy Parameters}
Dividing the self--energies by twice the (invariant) omega--energy $E_\omega=\sqrt{q^2 + m^2_\omega}$ Schr\"{o}dinger-like
complex--valued optical potentials are obtained:
\be
V^{(A)}_{L/T}(\mathbf{r},q)=\Sigma_{L/T}(\mathbf{r},q)/(2E_{\omega})=U^{(A)}_{L/T}(\mathbf{r},q)-iW^{(A)}_{L/T}(\mathbf{r},q).
\ee
In shape, their radial form factors are equal to those of the self--energies of Fig.\ref{fig:SrLT} at the same energy, albeit the magnitudes change. Meaningful measures, representing universal characteristics of potentials, are the radial moments
\be \label{eq:Moments}
M_{L/T}(q|n)=\int d^3r r^n V^{(A)}_{L/T}(\mathbf{r},q)=M^{(U)}_{L/T}(q|n)-iM^{(W)}_{L/T}(q|n).
\ee
Of particular interest are the moments $n=0$ and $n=2$ serving to define the volume integrals per nucleon $I_{L/T}$ and the mean--square radii $R_{L/T}$ of the potentials:
\bea
I_{L/T}(q)&=&M_{L/T}(q|0)/A=I^{(U)}_{L/T}(q)-iI^{(W)}_{L/T}(q)\\
R^{(U/W)2}_{L/T}(q)&=&\frac{M^{(U/W)}_{L/T}(q|2)}{M^{(U/W)}_{L/T}(q|0)}.
\eea

\begin{figure}[h]
\centering
%\sidecaption\overrightarrow{}
\includegraphics[width=7cm,clip]{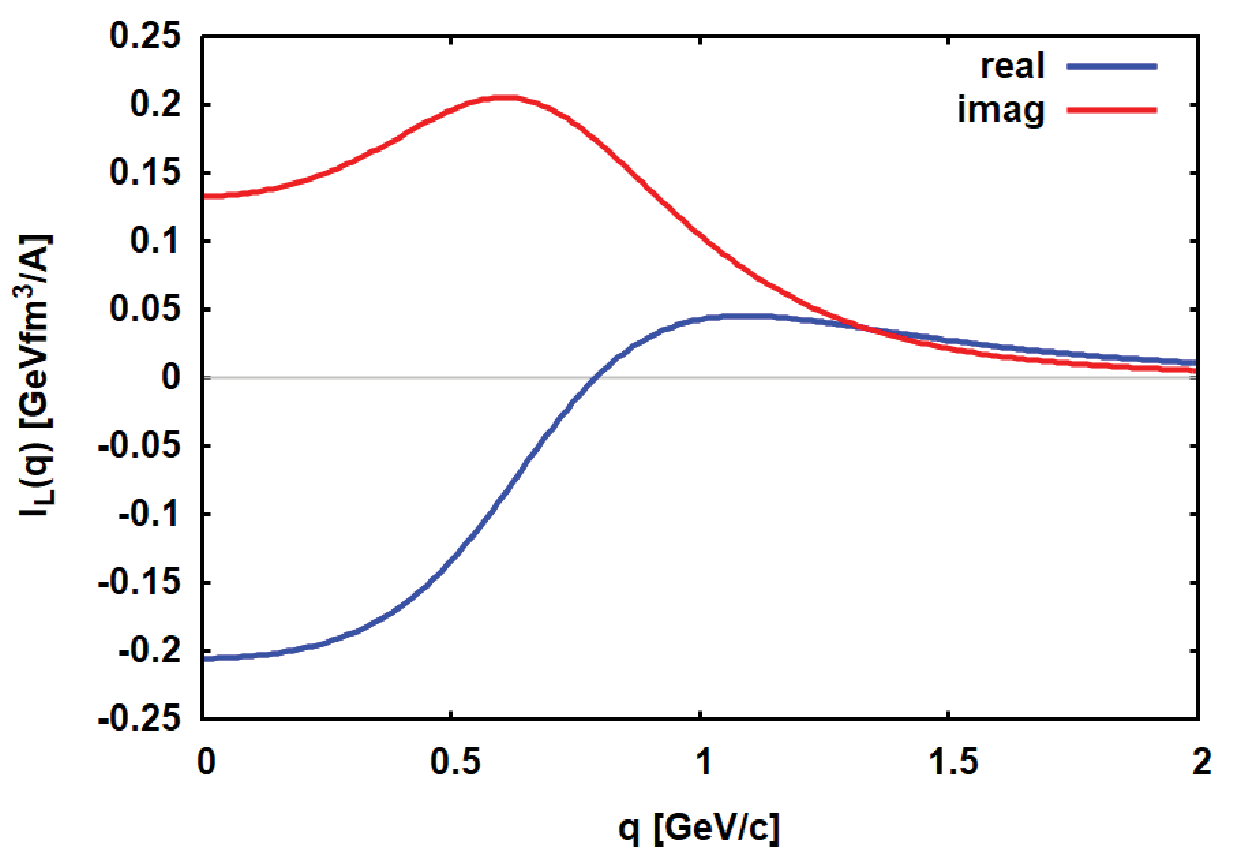}
\includegraphics[width=7cm,clip]{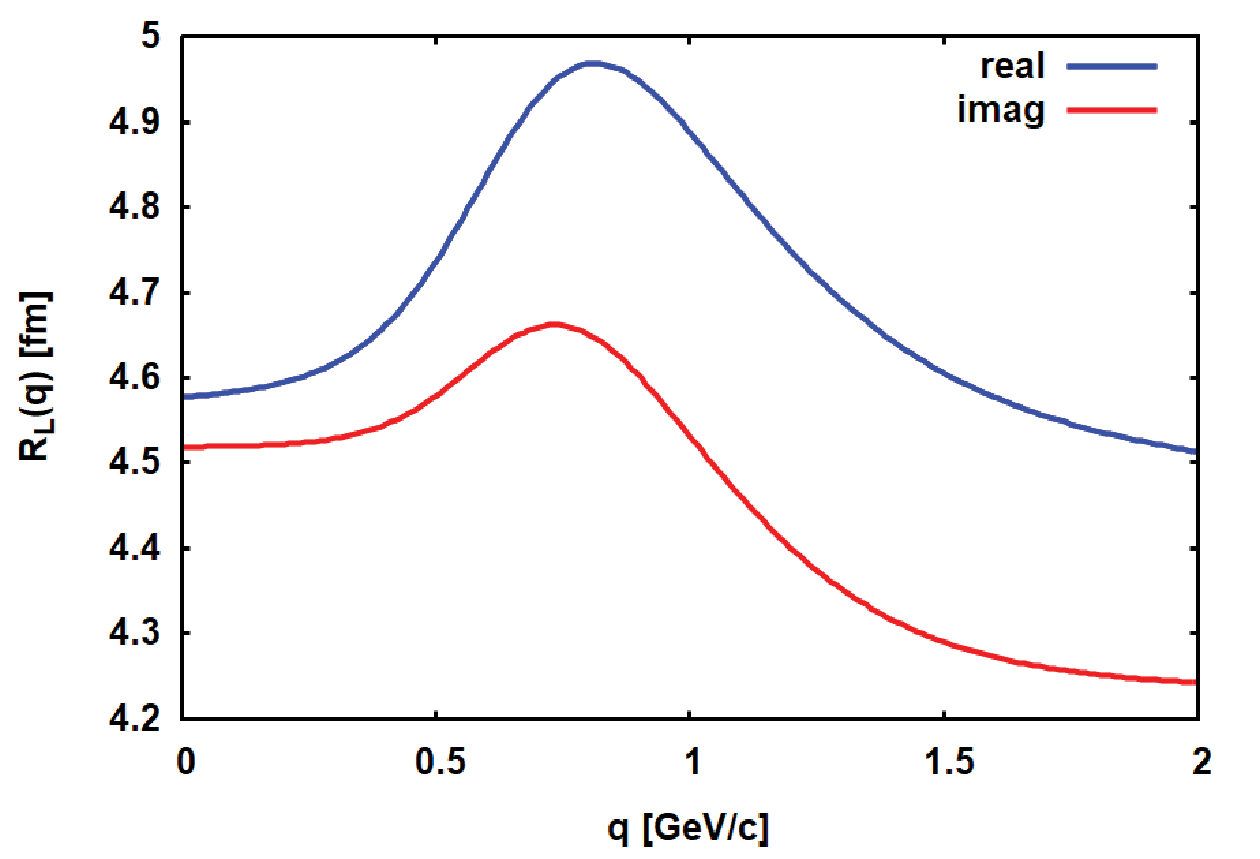}
\includegraphics[width=7cm,clip]{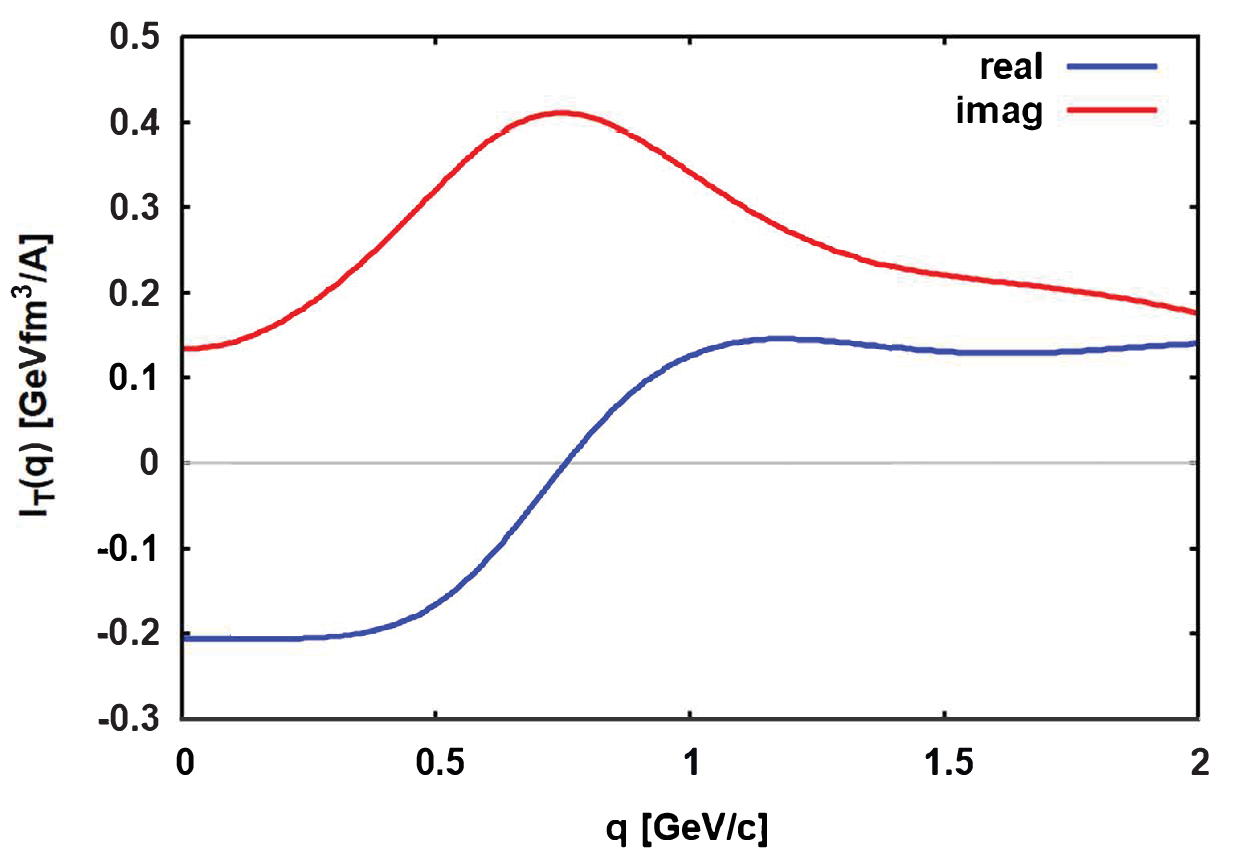}
\includegraphics[width=7cm,clip]{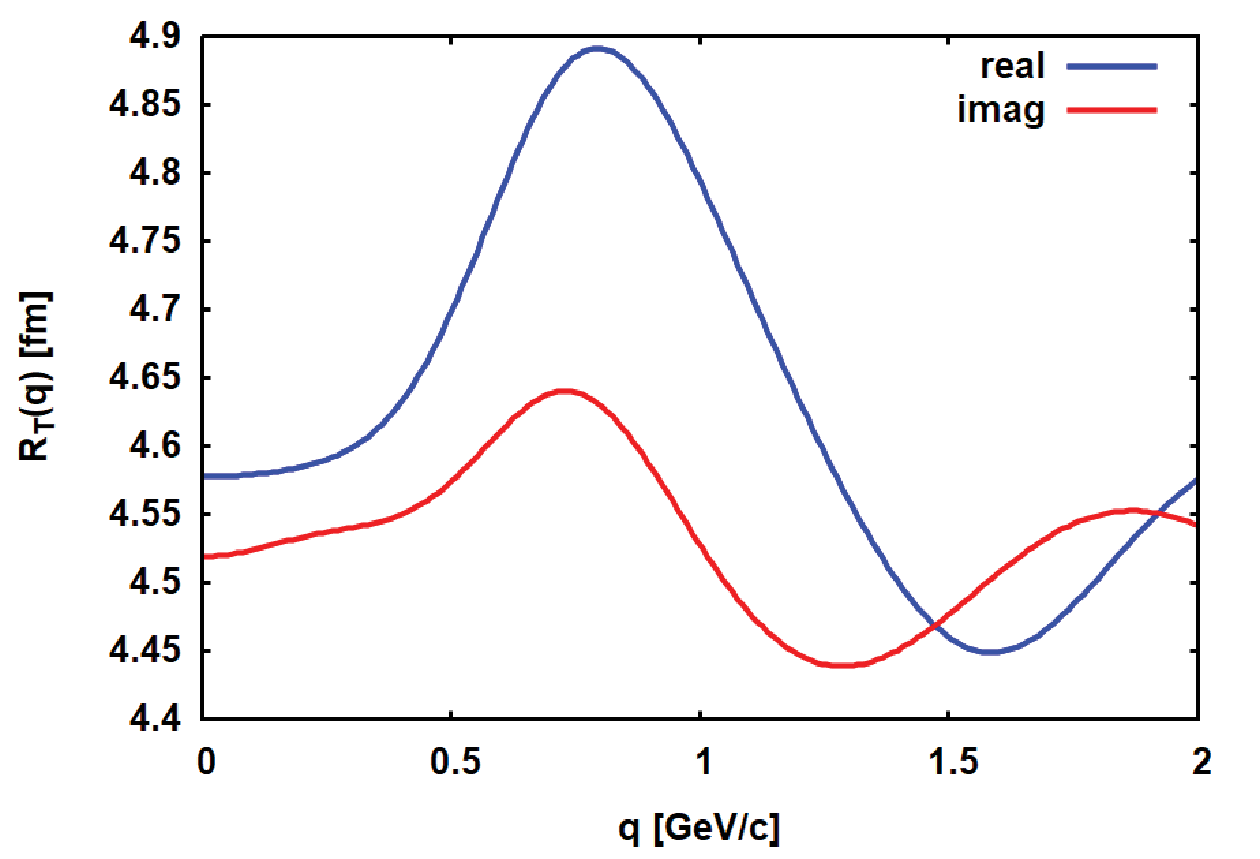}
\caption{Volume integral per nucleon (left) and rms--radii (right) of the Schr\"{o}dinger--like potentials for $\omega +{}^{93}$Nb. Results for the longitudinal (top row) and the transversal (bottom row) potentials, respectively, are shown. }
\label{fig:Pots}
\end{figure}

The volume integrals per particle represent the overall strength of the interactions. The root--mean--square (rms) radii $R^{(U)}_{L/T}$ and $R^{(W)}_{L/T}$ provide information on the geometry of the real and imaginary potentials.
In Fig.\ref{fig:Pots}, volume integrals and rms--radii of longitudinal and transversal potentials for $\omega + {}^{93}$Nb are displayed as functions of the invariant relative momentum $q=q(s)$, Eq. \eqref{eq:qcm}. The afore mentioned transition around $q\sim 0.8$~GeV/c of the real self--energies from attraction to repulsion is seen now very clearly. As typical for a dispersion functional connection between imaginary and real parts, the node of the real strength is located at (or close to) the momentum where the imaginary component acquires a maximum.

The moments are ideal quantities to compare interactions of different systems taking care of the correlation between potential form factors and depths. As known from decade--long studies of optical model potentials of strongly absorbing systems, none of those quantities by itself is meaningful but suffers from ambiguities, see e.g. \cite{Wong:1984fy,Brandan:1992kyu}. That problem is avoided by using potential moments which reflect universal properties of interacting systems. The varying shapes and sizes of the radial self--energies, Fig.\ref{fig:SrLT}, reflected also by the energy dependence of the rms--radii, Fig.\ref{fig:Pots}, underline the ambiguity problem clearly. Therefore, here we refrain from discussing potential depths separately as was done e.g. \cite{Friedrich:2016cms} and similar studies, frequently assuming direct proportionality of potentials to the nuclear density distribution at all energies.

Typically, the rms--radii of potentials follow the rms-radius of the ground state density, possibly augmented by the range of the projectile--target interaction. For $^{93}$Nb the RMF calculations lead to a total ground state density with $\sqrt{<r^2>}\simeq 4.59$~fm. However, the potential rms--radii of Fig.\ref{fig:Pots} rarely match that value: Only close to threshold the real parts attain rms--radii of that magnitude whereas the rms--radii of the imaginary potentials start at lower values. Around the transition from attraction to repulsion the sizes of the potentials increase by about 10\% reaching for the real parts values close to 5~fm while at higher energies the rms--radii decrease. In the transversal potentials the interference of S--wave and P--wave components lead to an oscillatory pattern.

\subsection{Low--energy Parameters}
Once self--energies and the deduced potentials are available, it is tempting to investigate global properties of $\omega$--nucleus dynamics. Of special interest is the behaviour of an interacting quantum system at low energies. Setting the focus on $\omega + A$ s--wave interactions, the dynamics at threshold are characterized by the scattering lengths $a_{L/T}$ and the effective ranges $r_{L/T}$, describing the dependence of the s--wave phase shifts $\delta_{L/T}(q)$ at the kinematical threshold $q\to 0$ according to the well known effective range formula
\be \label{eq:EffExp}
q\cot{\delta_{L/T}}\to -\frac{1}{a_{L/T}}+\frac{1}{2}q^2r_{L/T},
\ee
which corresponds to a next--to--lading order momentum expansion of the s--wave K-matrix. Close to threshold the longitudinal and transversal potential converge to the same limiting values. Hence, one finds for the scattering lengths $a_L=a_T=a_s$ and, correspondingly,  $r_L=r_T=r_s$ for the effective ranges. Since the potentials are complex--valued, also the derived scattering length $a_s=5.6542 -i0.9041$~fm and the effective range $r_s=3.7669 -i0.5759~fm$ are complex--valued.

\subsection{Indications for a $\omega+{}^{93}$Nb Bound State?}

Excitingly and unexpectedly, the real part of $a_s$ is positive. A positive scattering length, however, indicates the presence of a weakly bound $\omega + {}^{93}$Nb configuration. The most well known example of this kind is the triplet--even (TE) nucleon--nucleon interaction channel where the positive scattering ($a_{TE}\simeq 5,427$~fm, $r_{TE}\simeq 1.755$~fm) is reminiscent of the deuteron with binding energy $E_d=2.224$~MeV. In such cases, the effective range expansion, Eq.\eqref{eq:EffExp}, provides a decent extrapolation into the nearby below--threshold region, allowing to explore the bound state as pole of the S--matrix. In low--energy approximation, the poles are determined by the roots of $k\cot{\delta(k)}-ik=0$. Using Eq.\eqref{eq:EffExp}, the physical relevant solution is $k_B=i\left(1-\sqrt{1-2r_s/a_s}  \right)/r_s$, corresponding to an energy with real part $\varepsilon_B=-0.488$~MeV and an imaginary part corresponding to a dissipative width $\Gamma^{(A)}_B=4.445$~MeV. Adding the free decay width the total width is $\Gamma_B=13.125$~MeV. Hence, the model calculations predict a very weakly bound state close to threshold with a width of about 25 times the binding energy. Thus, a major part of the spectral distribution overlaps with the $\omega +{}^{93}$Nb continuum, implying a rapid decay of that states. One must be aware of the uncertainties on the existence and location of nuclear omega bound states because they will be affected in addition by couplings to the nuclear scalar and vector mean--fields and the afore mentioned higher order dissipative self--energies affecting the mesonic decay channels of the omega.

Although the above analysis does not exclude a tighter bound $\omega +{}^{93}$Nb s--wave state, it clearly rules out the deeply bound states appearing in other approaches. For example,
some time ago, Klingl et al. \cite{Klingl:1998zj} investigated omega bound states in finite nuclei, assuming at that time potentials as deep as -100~MeV at the nuclear center. That conjecture is not supported by the present results. With their model assumptions they obtained rather deeply bound omega s--wave states, e.g. in $^{39}$K a 1s--state with $\varepsilon_B \sim -87$~MeV while here no evidence is found for strong omega--binding in a nucleus.

\section{Discussion and Outlook}\label{sec:SumOut}
Interactions of omega mesons in nuclear matter and finite nuclei were investigated in a microscopic model based on dispersive self--energies including $NN^{-1}$ and $N^*N^{-1}$ modes. Different to other approaches,  self--consistent relativistic mean--fields (RMF) from relativistic Hartree theory were included for nucleons and resonances. The polarization tensors were calculated in asymmetric nuclear matter with the proton--to-neutron ratio of $^{93}$Nb. The coupling constants were determined directly by fits to the data of Ref. \cite{Friedrich:2016cms} on the omega in--medium width in $^{93}$Nb at central density. The data could be well described with dispersive self--energies of particle--hole type originating from P--wave and S--wave modes only. Initially, the complete spectrum of seven P--wave and seven S--wave resonances up to about 2.5GeV were used. Finally, however, in addition to the $NN^{-1}$ high-energy tail only two P--wave ($P_{11}(1710)$,$P_{11}(2300)$) and two S--wave ($S_{13}(1875)$,$S_{13}(2120)$) states were found to contribute significantly to $N^*N^{-1}$ polarization self--energies describing the observed width. The  S--wave polarization modes are essential for the proper description of self--energies in the threshold region while the modes involving $P_{11}$--states become important at higher omega--energies. However, the derived coupling constants are affected by considerable uncertainties of up to 40\%. The P-- and S--wave self--energies were projected into longitudinal and transversal self--energies, revealing the persisting dominance of transversal modes in nuclear matter.

The coupling of vector mesons to polarization modes leads naturally to transversal and longitudinal self--energies. The available data, however, do not distinguish such details which has to be accounted for by theory when adjusting parameters on measured values. The model calculations predict that the width distribution is dominated by about 90\% by the transversal in--medium self--energies, to which both P-wave and S--wave modes contribute. It is worth emphasizing that these findings imply that vector current conservation persists in nuclear matter still on a level of 90\%. The role played by the value of the width at threshold, $\Gamma_{thr}$, was investigates in exploratory numerical studies showing that $\Gamma_{thr}$ controls the spectroscopic composition of self--energies over the whole energy range.

The momentum dependent variations of up-- and downward shifts of the peak structures in the theoretical spectral functions point to a problem which requires special attention for the interpretation of spectroscopic data. The important, in fact \textsl{no--surprise}, message is that peak structures in mass distributions are affected by the self--energies where the momentum dependence of the real parts of self--energies is of special importance. As seen e.g. in Fig. \ref{fig:SwCSLT} the real parts change as functions of momentum from attraction to repulsion thus producing downward and upward shifts in spectral distributions. These shifts are pure many--body effects. They do not indicate chiral restoration.   The calculated spectral distributions strongly emphasize that first the in--medium dynamics of a meson has to be understood in due detail before safe conclusions on "chiral restoration" can be drawn.

The self--energies, calculated initially in infinite nuclear matter, were used in Local Density Approximation (LDA) to derive radius-dependent self--energies for the $\omega +{}^{93}$Nb system. The LDA self--energies were obtained by using the RMF--Hartree proton and neutron ground state densities, thus guaranteeing consistency between the spectroscopic content of self--energies and their radial distribution. In a further step, Schr\"{o}dinger--type potentials were defined. Volume integrals per nucleon and rms--radii of the real and imaginary potentials were studied as functions of the invariant $\omega +{}^{93}$Nb momentum. The considerable variations over the energy range cast doubts on the validity of the frequently used separability hypothesis of energy and density dependence.
The logical next step for quantitative studies of omega interactions on a finite nucleus is to formulate a reaction theory accounting properly for transversal and longitudinal meson--nucleus interactions. Such a theory should also account for omega--nucleus interactions of mean--field type obtained by coupling to the condensed nuclear scalar and vector fields. That requires the knowledge of three-meson vertices with tree--level coupling constants $f_{\omega\omega\sigma}$ and $f_{\omega\omega\omega}$. Hints on the values of these couplings may be derived from the SU(3) description of meson--meson--baryon interactions, see e.g. Ref.\cite{Ramos:2013mda}, but leaving the coupling constant largely undetermined.

The Schr\"{o}dinger--like potentials have been used to determine the $\omega+{}^{93}$Nb low--energy parameters. Solving the Schr\"{o}dinger equation at threshold energies, the s--wave K-matrix was calculated and analyzed in the effective range expansion up to order $q^2$. The positive scattering length signaled the existence of a $\omega + {}^{93}$Nb bound state just below threshold. The large width, however, indicates that about 50\% of the spectral strength is shifted into the continuum region, making the state partially particle--unstable.
The results of Sect. 4.4 and Sect. 4.5 are encouraging for the ongoing searches for bound states of heavy mesons. However, further studies are necessary addressing the full spectrum of $\omega + A$ interactions from static mean--field potentials to dynamical self--energies. Because of the long life time of the omega--meson, $\omega +A$ bound states are the ideal cases for detailed studies of static and dynamical meson--nucleus interaction.

Last but not least, it is worth to return to Fig. \ref{fig:Gw_Expl} where the results of the separate P-- and S--wave fits are displayed. The S--wave results are especially instructive for a comparison to the large threshold widths predicted by former meson cloud studies, amounting to $\Gamma_{thr}=121\pm 10$~MeV by Ramos et al. \cite{Ramos:2013mda} and to $\Gamma_{thr}=130\ldots 200$~MeV by Cabrera and Rapp \cite{Cabrera:2013zga}. As discussed in the context of Fig. \ref{fig:Gw_Expl} depending on the fitting strategy the large error bars of data close to threshold induce a corresponding large spread of results, finally leading to $\Gamma_{thr}= 50^{+204}_{-17}$~MeV. Thus, within the error bars the present approach leads to results fully compatible with those of the meson cloud calculations although at first sight the assumed physical scenarios seem to be quite different. Interestingly, the S--wave width distribution is of pure $N^*$ character, thus complying with the arguments given in App. \ref{app:Mapping} because the S--wave resonances contain naturally a $\pi + N$ component and even multi--pion configurations.

Here, interactions in cold nuclear matter at normal density and under mechanical and thermodynamical equilibrium conditions were considered. That state is hardly comparable to the environment encountered in high energy heavy ion central collisions as explored e.g. in \cite{Rapp:2009yu}. An important question left for future work is to clarify if there may be conditions under which leading the particle--hole contributions and the dynamically more complex sub--leading meson cloud mechanisms will be distinguishable.

\section*{Acknowledgements}
Inspiring and illuminating discussion with Mariana Nanova and Volker Metag are gratefully acknowledged. This work was supported in part by DFG grant Le439/16.
\newpage

\appendix

\section{Self-Energy Ambiguities}\label{app:Mapping}

\begin{figure}[h]
\centering
%\sidecaption
\includegraphics[width=6cm,clip]{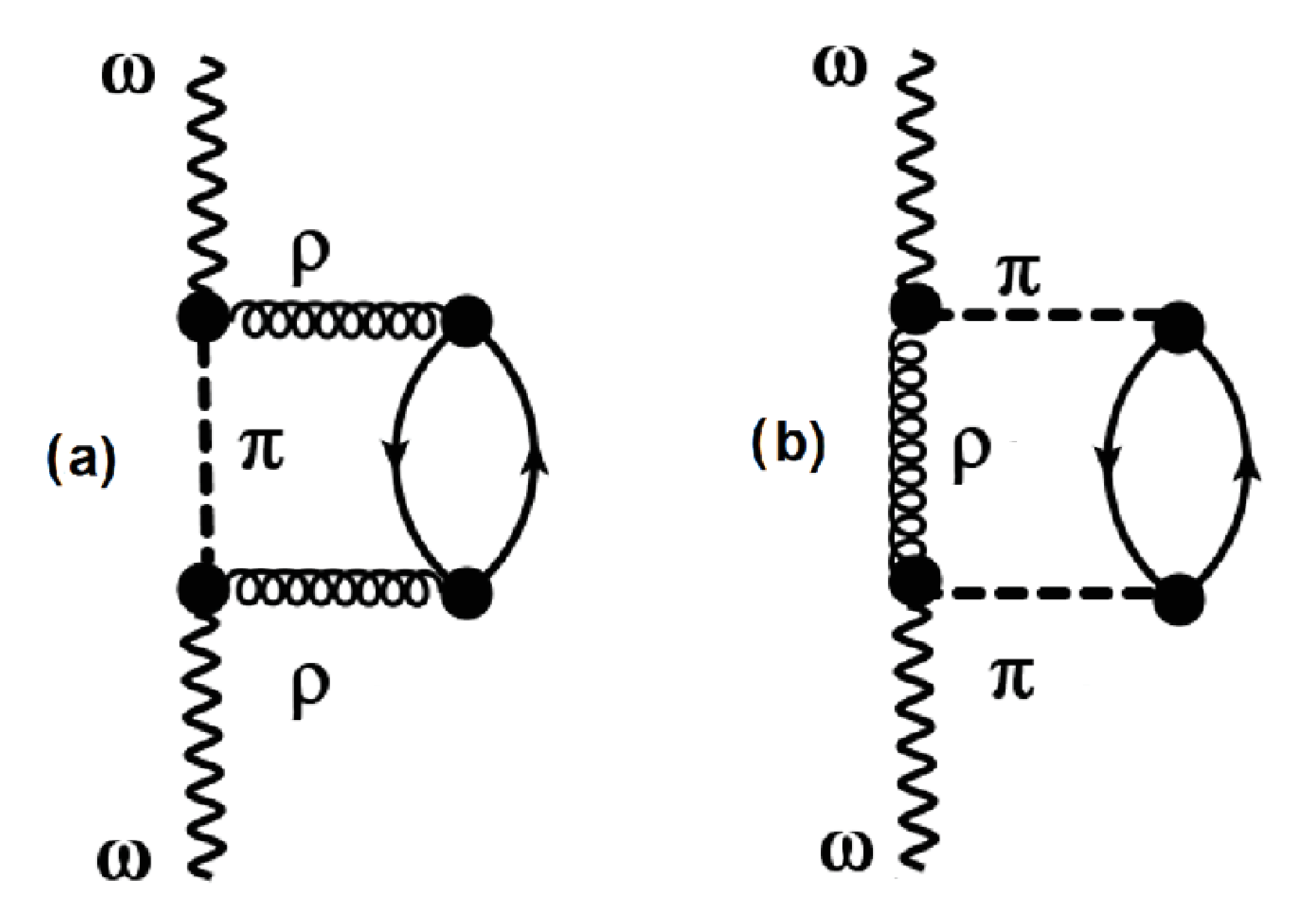}
\caption{In-medium omega--meson interactions of higher order involving meson loops modifying the mesonic decay channels illustrated by the case $\omega\to \rho\pi$. Graph $(a)$ describes
the coupling of the intermediate rho--meson and graph $(b)$ the coupling of the pion to dispersive self--energies as in Fig.\ref{fig:Graphs}, including both $NN^{-1}$ and $N^*N^{-1}$ modes of the background medium.}
\label{fig:wpirho}
\end{figure}

In this appendix the self--energy ambiguity mentioned in the introduction is elucidated qualitatively for the dominating $\omega\to \pi + \rho$ decay channel. In Fig.\ref{fig:wpirho} Born diagrams are shown which contribute to the coupling of either the pion or the rho-meson to (isovector) particle--hole excitations of the background medium. In such processes -- studied extensively by various theoretical methods e.g. in \cite{Klingl:1997kf,Lutz:2001mi,Muehlich:2006nn,Rapp:2009yu,Ramos:2013mda,Cabrera:2013zga} -- the decay and the propagation of the emitted mesons are modified by dispersive self--energies as of Fig.\ref{fig:Graphs}. However, although the present data are of outstanding high value, they provide information only on $\Gamma^{(A)}_{tot}$ but not on partial in--medium decay widths. Hence, branching ratios e.g. to the in free space dominating $\pi + \rho$ channel are not available. The lack of detailed information on the decay spectroscopy poses problems for theoretical model building. While in free space, the source of emitted mesons (or other radiation) is easy to identify this is no longer true in a nuclear environment. Mesons can result either from the direct decay of the omega, e.g. $\omega\to\pi+\rho\to 3\pi$, or they can be produced in $\omega +N\to N^*\to \pi+N$  processes, including at high enough energies even multi--pion emission.

\begin{figure}[h]
\centering
%\sidecaption
\includegraphics[width=6cm,clip]{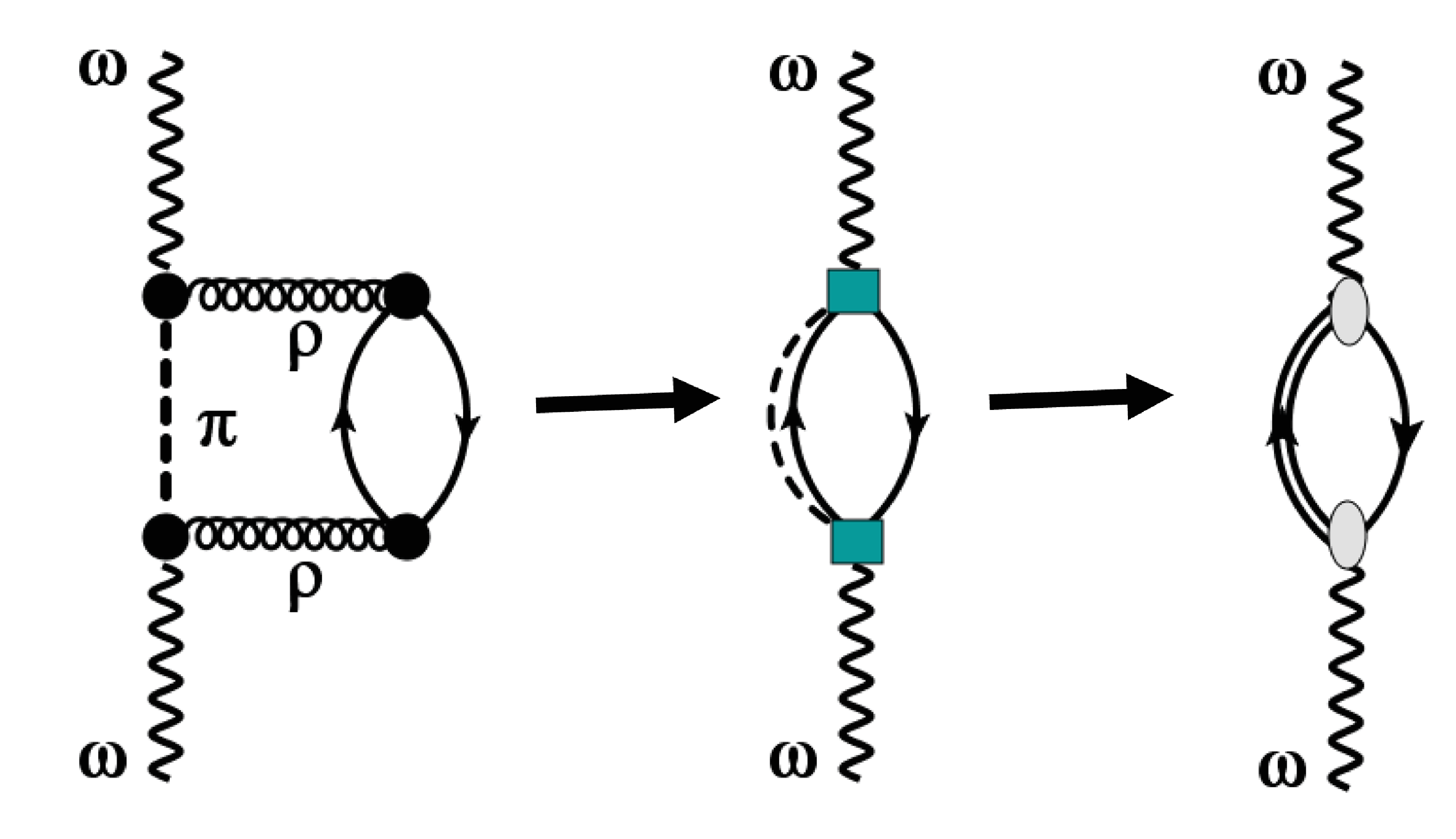}
\caption{Illustrating the relation of in--medium meson loops to baryon particle-hole loops for the $\omega \to \pi +\rho$ decay for the case where the rho-meson couples to an isovector excitation of the nuclear background medium.  }
\label{fig:Mapping}
\end{figure}

As illustrated in Fig. \ref{fig:Mapping} the in--medium meson cloud scenario and the particle--hole polarization scenario are overlapping and their contributions to the omega in--medium width are hardly distinguishable. Hence, ambiguities occur because part of the in--medium meson--loop contributions will be already be contained in the (leading) dispersive particle-hole self-energies, especially those given by $N^*N^{-1}$ excitations.

As an example, we consider the case $\omega\to \pi +\rho$ where he rho--meson induces a nuclear particle-hole mode (left diagram) the pion moves freely. The 4--momentum brought in by the omega--meson is divided among the decay particles taking care of by an appropriate integration over intermediate momenta. The self--energy is obtained by de--excitation of the background mode and recombination $\pi+\rho \to \omega$. For momentum transfers less than the meson rest mass the rho--meson propagator may be replaced by a constant, leading to an effective point-coupling interaction as depicted in the center diagram. The pion may be attached to the $N$ or $N^*$ particle state, thus contributing to a nucleon resonance configuration, displayed in the right hand graph.

Apparently, one faces the principal problem that there is an overlap between the omega self--energies induced by the direct coupling to $NN^{-1}$ and $N^*N^{-1}$ modes, Fig. \ref{fig:Graphs}, and the (sub--leading) self--energies affecting the meson cloud produced by the decaying omega. The same kind of result is obtained by exchanging the role of the pion and the rho--meson. More exotic (but less important) channels e.g. involving combinations of strange and anti-strange mesons like $K\bar{K}$ channels as considered e.g. in Ref. \cite{Ramos:2013mda} or effects related to in--medium $\omega$--$\phi$ mixing may be treated similarly.

Obviously, inclusive $\Gamma_{\omega A}$ data alone are not sufficient to distinguish the contributions from the competing channels like the one discussed here. Moreover, including straight forwardly both the leading polarization self--energies and sub--leading in--medium meson cloud self--energies in an approach relying on empirical input will lead to overcounting.

\section{The Vector--Meson Propagator and Spectral Functions}\label{app:freeVM}

The propagator of a free, stable spin-1 particle with bare mass $m_0$ and four--momentum $q=(q_0,\mathbf{q})^T$ is given by:
\begin{eqnarray}
\label{eq:freeprop}
D^{0 \,\mu \nu}(q) &=& \frac{-(g^{\mu \nu}-\frac{q^{\mu}\,q^{\nu}}{m_0^2})}
                            {q^2-m_0^2} \nonumber\\
 &=& \frac{-(g^{\mu \nu}-\frac{q^{\mu}\,q^{\nu}}{q^2})}
                            {q^2-m_0^2} +
                                \frac{1}{m_0^2}\,\frac{q^{\mu}\,q^{\nu}}{q^2}
\quad ,
\end{eqnarray}
where in the second line of Eq.~(\ref{eq:freeprop}) we have separated the
propagator into a longitudinal and a transversal part.  The coupling of
the meson to (pionic) decay channels gives rise to a self-energy
$\Sigma_{v}^{\mu \nu}(q^2)$. Current conservation demands that the
self--energy is transversal, expressed by:
\begin{equation}
\label{eq:currcons}
        q_{\mu}\,\Sigma_{v}^{\mu \nu}(q^2) = 0
\quad .
\end{equation}
Because of this, only the first term in Eq.~(\ref{eq:freeprop}) is
modified, when a self-energy is taken into account:
\begin{eqnarray}
\label{eq:vacprop}
D^{\mu \nu}(q)  &=& \frac{-(g^{\mu \nu}-\frac{q^{\mu}\,q^{\nu}}{q^2})}
                            {q^2-m_0^2 - \Sigma_{v}(q^2)} +
                                \frac{1}{m_0^2}\,\frac{q^{\mu}\,q^{\nu}}{q^2}
\end{eqnarray}
The explicit form and composition of $\Sigma_{v}$ depends on the vector meson under consideration, e.g. Ref. \cite{Peters:1997va} for the rho--meson case.

The real part of $\Sigma_{v}$ is taken into account approximately
by the introducing the physical mass $m_v\sim m_0+Re(\Sigma_{v}^{00}(q^2_v))$
where $q^2_v\equiv m^2_v$.
Thus we use:
\begin{eqnarray}
\Sigma_{v}(q) \equiv - {\rm i} \;m_v \;\Gamma_{v}(q^2)
\quad .
\end{eqnarray}
For two--body decay channels $v\to m_1m_2$ of angular momentum $\ell$ the total decay width  is appropriately
parameterized as:
\begin{eqnarray}
\label{eq:vwidth}
        \Gamma_{m_1m_2}(m^2) = \Gamma_0 \,\frac{m_v}{m}\,
        \left(\frac{q\left(m\right)}{q\left(m_v \right)}\right)^{2\ell+1} \,\,.
\end{eqnarray}
$q(m)=|\mathbf{q}(m)|$ is the three--momentum of the emerging cloud of decay particles in the rest frame of the decaying
vector meson of invariant mass $m_v$; $\Gamma_0$ is
the width at rest.  Eq.~(\ref{eq:vwidth}) contains precisely the
energy dependence of the imaginary part of the self-energy that comes
out of a one loop calculation. At  high invariant mass $m\gg m_v$, regularization by an appropriately chosen
form factor will be necessary.

A well studied example is the mentioned 2--pion p--wave decay of the rho--meson, e.g. \cite{Peters:1997va}. A different situation is encounter for the omega decays ending finally in three--pion compound, see \cite{Niecknig:2012sj} for a dispersion theoretical description.
As is obvious from the rather tiny total width of the omega--meson, $\Gamma_{0}=8.68$~MeV, the mesonic decay channels are much less important for an omega--meson than for a rho--meson.
The proper description of three--pion decay of the omega--meson is more involved but may be treated effectively
by the dominant $\omega\to\rho + \pi \to 2\pi +\pi$ process depicted in Fig.\ref{fig:wpirho}. However, the broad spectral distribution
of the intermediate rho--meson must be taken into account -- as discussed e.g. in \cite{Muehlich:2006nn}.

The most important contribution of real part of the meson--loop self--energy is taken care of by using the physical meson mass, $m^2_v=m^2_0+Re(\Sigma_v(m_v))$. Then, the energy dependence of the residual self--energy $\widetilde{\Sigma}_v(m^2)=\Sigma_v(m^2)-Re(\Sigma_v(m^2_v))$ has
only little influence on the vacuum spectral function defined as:
\begin{eqnarray}
\label{eq:freespec}
A_v(q^2) = -\frac{1}{\pi} \; {\rm Im}\left(
        \frac{1}{q^2 - m_0^2 -\Sigma_{v}}\right)=
 -\frac{1}{\pi} \; {\rm Im}\left(
        \frac{1}{q^2 - m_v^2 -\widetilde{\Sigma}_{v}}\right)
\quad
\end{eqnarray}
and is therefore neglected here.

In nuclear matter, Lorentz-invariance is broken and vector meson self-energies are
characterized by two scalar functions instead of one. Using the
transversal projector:
\begin{eqnarray}
\label{eq:projectorst}
P_T^{\mu\nu} &=& - \left(
        \begin{array}{cc}
                0&0\\
                0&\delta_{ij} - \frac{q_iq_j}{\mathbf{q}\,^2}\\
                \end{array}
                \right)
\end{eqnarray}
and the longitudinal projector, respectively:
\begin{eqnarray}
\label{eq:projectorsl}
P_L^{\mu\nu} &=& (g^{\mu \nu}-\frac{q^{\mu}\,q^{\nu}}{q^2}) - P_T^{\mu\nu}
\nonumber \\
&=& -\left(
        \begin{array}{cc}
          \frac{\mathbf{q}\,^2}{q^2}&\frac{\omega q_j}{q^2}\\
          \frac{\omega q_i}{q^2}&\frac{ \omega^2 q_iq_j}{\mathbf{q}\,^2 q^2}\\
                \end{array}
                \right)
\end{eqnarray}
the in--medium self-energy is decomposed
\begin{eqnarray}
        \Sigma^{\mu\nu}(q)=
                P_L^{\mu\nu}\,\Sigma_L(q) +
                P_T^{\mu\nu}\,\Sigma_T(q)
\end{eqnarray}
into the longitudinal and transversal parts
\begin{eqnarray}
\begin{array}{rcrl}
        \Sigma_L(\omega,\mathbf{q}) &=&
                & P_L^{\mu\nu}\Sigma_{\mu\nu}(\omega,\mathbf{q})
\\
\\
        \Sigma_T(\omega,\mathbf{q}) &=& \frac{1}{2} &
                P_T^{\mu\nu}\Sigma_{\mu\nu}(\omega,\mathbf{q})
\quad .
\end{array}
\end{eqnarray}
Then the dressed vector meson propagator becomes:
\begin{eqnarray}
        D^{\mu\nu}(w,\mathbf{q})=-\frac{P_L^{\mu\nu}}
                {q^2-m_\rho^2-\Sigma_L(w,\mathbf{q})-\Sigma_v(w,\mathbf{q})}
                -\frac{P_T^{\mu\nu}}{q^2-m_\rho^2-\Sigma_T(w,\mathbf{q})-\Sigma_v(w,\mathbf{q})}
                +\frac{q^\mu q^\nu}{m_v^2 q^2}
\quad .
\end{eqnarray}
Thus, a transversal and a longitudinal spectral function can be defined:
\begin{eqnarray}
\label{eq:fullspec}
A_{T/L}(w,\mathbf{q}) = - \frac{1}{\pi} \;{\rm Im}\left(
        \frac{1}{w^2-\mathbf{q}^2 - m_v^2 -\Sigma_{T/L}(w,\mathbf{q})
        -\Sigma_{v}(w,\mathbf{q})}
\right) \quad .
\end{eqnarray}
For $\mathbf{q}=0$,  $A^T=A^L$ is obtained.

\section{$\omega$--Baryon Interactions}

For the present study the dynamics are described in non-relativistic reduction, still obeying the global intrinsic symmetries, including current conservation. For four--momentum transfer $q^\mu=(\nu,\mathbf{q})^T$ and the omega--meson given by the four--vector $\omega^\mu=(\omega^0,\bm{\omega})^T$ the interactions are defined by the set of Lagrangians
\begin{eqnarray}
\label{eq:Lagrange}
\begin{array}{llllll}
{\cal L}^{(J)}_{int}&=&       \frac{g_{N R/N \omega}}{m_\omega}\,\,
                \psi_{R/N}^\dagger\left(\bm{\sigma}\times \mathbf{q}\right)\cdot\bm{\omega}\,\psi_N
                &\mbox{for}&  J = \frac{1}{2}^+                            \\
&=&     \frac{g_{N R\omega}}{m_\omega}\,\,
                \psi_{R}^\dagger\,\left(\mathbf{S}\times \mathbf{q}\right)\cdot\bm{\omega}\,\psi_N
                &\mbox{for}&  J = \frac{3}{2}^+                            \\
&=&     \frac{g_{N R\omega}}{m_\omega}\,\,
                \psi_{R}^\dagger\,\left(\bm{\sigma}\cdot\bm{\omega}\,\nu
                - \omega^0\bm{\sigma}\cdot\mathbf{q}\right) \,\psi_N
                &\mbox{for}&  J = \frac{1}{2}^-                            \\
&=&     \frac{g_{N R\omega}}{m_\omega}\,\,
                \psi_{R}^\dagger\,\left(\mathbf{S}\cdot\bm{\omega}\,\nu
                - \omega^0 \mathbf{S}\cdot \mathbf{q} \right)\,\psi_N
                &\mbox{for}& J =  \frac{3}{2}^-
\quad .
\end{array}
\end{eqnarray}
$\mathbf{S}$ is the spin--$\frac{1}{2}\to \frac{3}{2}$ transition spin operator, nucleon and resonance field operators are denoted by $\psi_N$ and $\psi_R$, respectively. The $J=\frac{1}{2}^+$ case includes the coupling to the nucleon, as indicated.
Current conservation of the elementary vertices -- as incorporated above -- is an essential prerequisite
that the omega in--medium self--energies fulfill that condition as well \cite{Peters:1997va}.
The coupling constants are determined directly from data by a fitting procedure discussed in the main text.

\section{Self--Energies of an $\omega$--Meson in Nuclear Matter}\label{app:SelfPS}

\subsection{Nuclear Polarization Tensors}\label{sapp:PolProp}

Assuming that the interactions intrinsic to $A$ can be separated into (a matrix of) mean--fields $\mathcal{U}_A$ and (a matrix of) residual interactions $\mathcal{V}_A$ the polarization propagator is given formally by the Dyson equation
\bea
&&\Pi^{(MF)}_A=\Pi^{(0)}_A+\Pi^{(0)}_A\mathcal{U}_A\Pi^{(MF)}_A\\
&&\Pi_A=\Pi^{(MF)}_A+\Pi^{(MF)}_A\mathcal{V}_A\Pi_A .
\eea
$\Pi^{(0)}_A$ is the nuclear polarization propagator without interactions. With the susceptibility tensors
$\chi^{-1}_{MF}=1+\Pi^{(0)}_A\mathcal{U}_A$ and $\chi^{-1}_{A}=1+\Pi^{(MF)}_A\mathcal{V}_A$
the polarization propagators attain the structure
\bea
&&\Pi^{(MF)}_A=\chi_{MF}\Pi^{(0)}_A\\
&&\Pi_A=\chi_{A}\Pi^{(MF)}_A .
\eea
Thus, a meaningful leading order approach is to evaluate the polarization tenors in mean--field approximation, $\Pi^{\mu\nu}_A\approx \Pi^{\mu\nu}_{MF}$. In the following, we use $\Pi^{\mu\nu}\equiv \Pi^{\mu\nu}_A \equiv \Pi^{(MF)\mu\nu}_A$.

\subsection{Particle-Hole Lindhard Function}\label{app:Lindhard}
Particle--hole dynamics is contained in the reduced 4--point propagators, given here in a non--relativistic hybrid formulation by the Lindhard function:
\bea\label{eq:LindhardNB}
&&L_{NB}(w,\mathbf{q})=C_{NB}\int \frac{d^3k}{(2\pi)^3}\\
&&\left(
\frac{\theta(k^2_{F_N}-\mathbf{k}^2)\theta((\mathbf{k}+\mathbf{q})^2-k^2_{F_B})}{w+E^*_N(\mathbf{k})
-E^*_{B}(\mathbf{k}+\mathbf{q})+\Delta_{NB}+\frac{i}{2}\Gamma_{NB}}
+
\frac{\theta(k^2_{F_N}-(\mathbf{k}-\mathbf{q})^2)\theta(\mathbf{k}^2-k^2_{F_B})}{-w+E^*_N(\mathbf{k}-\mathbf{q})
-E^*_{B}(\mathbf{k})+\Delta_{NB}-\frac{i}{2}\Gamma_{NB}}
\right)\nonumber ,
\eea
with the spin--isospin factors $C_{NB}$ and the nucleon Fermi--momenta $k_{F_N}$, $N=p,n$.

The Lindhard functions by themselves behave for large spatial momenta asymptotically as $\sim \frac{1}{\mathbf{q}^2}$, which, however, is compensated by the transversal pre-factor $\varepsilon^{(B)}_T=\frac{\mathbf{q}^2}{m^2_v\mathbf{}}$, thus resulting in an constant value, growing in principle unlimited with the number of $N^*N^{-1}$ self--energies. In order to cure that problem \footnote{At first sight, a not fully satisfactory aspect of this solution is that the configuration space problem is shifted to the momentum dependence. However, at the mass--shell, the momentum cut.off is identical to an energy cut-off, thus regularizing the increasing level density. }, convergence is enforced by the (covariant)) cut--off vertex form factor
\be
\widetilde{F}_P(q^2)=\left(\frac{\Lambda^2_P}{\Lambda^2_P+q^2}  \right)^n ,
\ee
where $q$ is the invariant relative 3--momentum in the center--of--momentum frame:
\be\label{eq:qcm}
q^2=\frac{1}{4s}(s-(m_\omega+M_A)^2)(s-(m_\omega-M_A)^2)
\ee
determined by the invariant center--of--mass energy $s$.
Note that the form factor is equivalent to a cut--off in the resonance on--shell energies, $E^*_B(\mathbf{q})$.
In practice, dipole form factors (n=2) are used with cut--off $\Lambda_P=2$~GeV/c.

In addition to the mean--fields, there are energy--dependent dispersive self--energies $S_{NB}=D_{NB}-\frac{i}{2}\Gamma_{NB}$ accounting for dissipative effects of the nucleon hole states and the particle states $B=N,N^*$. Real and imaginary parts are connected up to a substraction constant $D_0$ through a dispersion relation. Hence, specifying the width is sufficient to determine the spectroscopy of an particle--hole configuration $BN^{-1}$. Here, we are interested only on a schematic description on a global level. A convenient approach, inspired by \cite{Mahaux:1982eig,BAKER:1997235}, is to use
\bea
&&\Gamma_{NB}(w,\rho)=\Gamma_{B}(w)+\Gamma_{N}(w)-\gamma_{NB}(w,\rho)\\
&&D_{NB}(w,\rho)=-\frac{P}{2\pi}\int^\infty_0dw'\frac{\Gamma_{NB}(w',\rho)}{w'^2-w^2}+D_0 .
\eea
The $NN^{-1}$ and $N^*N^{-1}$ particle--hole widths $\Gamma_{NB}$ are expressed in terms of the total free--space decay width $\Gamma_B$, the nucleon hole width $\Gamma_N$ and the (imaginary part of the) particle--hole correlation self--energy $\gamma_{NB}$, the latter accounting also for dependencies on the nuclear density $\rho$. Taking into account also  time--like vector mean--field potentials $U_{N,B}$, the $NB$--mass shift due to time--like self--energies is obtained as
\be
\Delta_{NB}(w,\rho)=U_N(\rho)-U_B(\rho)+D_{NB}(w,\rho).
\ee

\subsection{Nuclear Polarization Tensors and Self--Energies}

\paragraph{\textbf{Self--energies containing P--wave Resonances:}}

The coupling of a vector meson to P--wave resonances
carrying positive parity leads to transversal, current conserving self--energies of purely spatial character
\be
\Sigma^{\mu\nu}_{NB}(w,{\mathbf{q}})=(1-\delta^{\mu 0})(1-\delta^{\nu 0})\Sigma^{ij}_{NB}(w,{\mathbf{q}}):
\ee
Hence, $\Sigma^{\mu 0}\; = \;\Sigma^{0 \nu}\; = \; 0$ for $\mu,\nu=0\ldots 3$. The space--like components are given by:
\begin{equation}
\label{eq:SelfP}
        \Sigma^{ij}_{NB}(w,{\mathbf{q}}) =
        \frac{1}{2}\left(\delta^{ij}-\frac{q^i\,q^j}{\mathbf{q}\,^2}\right)\Sigma^{(P)}_{NB}(w,\mathbf{q}).
\end{equation}
where the reduced P--wave self--energy is given by
\be
\Sigma^{(P)}_{NB}(w,\mathbf{q})=2g^2_{NB\omega}F_P(\mathbf{q})\frac{\mathbf{q}^2}{m^2_\omega}
         L_{NB}(w,{\mathbf{q}}).
\ee
such that the trace becomes
\be
\sum_i\Sigma^{ii}_{NB}(w,{\mathbf{q}}) =\Sigma^{(P)}_{NB}(w,\mathbf{q}).
\ee
The total transversal P--wave self--energy is obtained by summation over nucleons and resonances:
\be\label{eq:TotalP}
\Sigma^{(P)}_T(w,\mathbf{q})=\sum_{N=n,p;B=N,N^*}\Sigma^{(P)}_{NB}(w,\mathbf{q}).
\ee
We define a generalized off--shell width distribution
\be
Im\left(\Sigma^{(P)}(w,\mathbf{q}) \right)=-\sqrt{q^2}\widetilde{\Gamma}^{(P)}_{\omega A}(w,\mathbf{q}).
\ee
For on--shell conditions, $q^2=m^2_\omega$, the P--state contribution to the physical width is obtained:
\be\label{eq:TotalGwP}
\Gamma^{(P)}_{\omega A}(\mathbf{q})=-\frac{1}{m_\omega}Im\left(\Sigma^{(P)}(E(\mathbf{q}),\mathbf{q}) \right),
\ee
which by Eq.\eqref{eq:TotalP} is seen to be a superposition of partial widths:
\be
\Gamma^{(P)}_{\omega A}(\mathbf{q})=\sum_{N,B}\Gamma^{(P)}_{NB}(\mathbf{q}).
\ee
Each partial width is defined analogously to Eq.\eqref{eq:TotalGwP}.

\paragraph{\textbf{Self--energies containing S--wave Resonances:}}

The negative parity
S--wave resonances, observed in $\pi N$ S--wave scattering, lead to  self--energies of a more complex tensorial structure:
\begin{equation}
\label{selfneg}
        \Sigma^{\mu\nu}_{NB}(w,\mathbf{q}) =
        T^{\mu\nu}\frac{1}{m^2_v}\Sigma^{(S)}_{NB}(w,\mathbf{q})
\quad .
\end{equation}
with the reduced S--wave self--energy
\be
\Sigma^{(S)}_{NB}(w,\mathbf{q})= g^2_{NB \omega}
        F_S(\mathbf{q}^2)L_{NB}(w,\mathbf{q})
\ee
Regularization is introduced by the cut--off form factor, defined as above:
\be
F_S(\mathbf{q})=\left(\frac{\Lambda^2_S}{\Lambda^2_S+m^2_v-q^2}  \right)^n
               =\left(\frac{\Lambda^2_S}{\Lambda^2_S+\mathbf{q}^2}  \right)^n.
\ee
In practice, as in the P--wave case a dipole form factor (n=2) is used with cut--off $\Lambda_S=\Lambda_P=2$~GeV/c.

The tensor
\begin{equation}
T^{\mu\nu}(w,\mathbf{q})=
        \left(
          \begin{array}{cc}
            {\mathbf{q}\,^2}&{w q_j}\\
            {w q_i}&w^2 \delta_{ij}\\
          \end{array}
        \right)
\quad ,
\end{equation}
is decomposed by means of Eq.\eqref{eq:projectorsl} into transversal ($P_T$) and longitudinal ($P_L$) components:
\begin{equation}
\label{decs}\mathbf{}
T^{\mu\nu}= w^2 P_T^{\mu\nu} + q^2 P_L^{\mu\nu}
\quad .
\end{equation}
Hence, the self--energies involving S-wave resonances are superpositions of longitudinal and transversal components:
\be
\Sigma^{\mu\nu}_{NB}(w,\mathbf{q})=P_T^{\mu\nu}\Sigma^{(T)}_{NB}(w,\mathbf{q})+P_L^{\mu\nu}\Sigma^{(L)}_{NB}(w,\mathbf{q})
\ee
where
\bea
\Sigma^{(L)}_{NB}(w,\mathbf{q})&=&\frac{q^2}{m^2_v}\Sigma^{(S)}_{NB}(w,q)\\
\Sigma^{(T)}_{NB}(w,\mathbf{q})&=&\frac{w^2}{m^2_v}\Sigma^{(S)}_{NB}(w,q).
\eea
Under on-shell conditions, $q^2=m^2_v$ and $w^2=\mathbf{q}^2+m^2_v$, one finds
\bea
\Sigma^{(L)}_{NB}(w,\mathbf{q})&=&\Sigma^{(S)}_{NB}(w,q)\\
\Sigma^{(T)}_{NB}(w,\mathbf{q})&=&\left(1+\frac{\mathbf{q}^2}{m^2_v}\right)\Sigma^{(S)}_{NB}(w,q).
\eea

%\bibliographystyle{unsrt}
%\bibliography{Meson}

\end{document}